\documentclass[aps,prc,twocolumn,amsmath,amssymb,showpacs,floatfix]{revtex4}
\usepackage{graphicx}
\usepackage{hyperref}

\def\lsim{\mathrel{\rlap{
\lower4pt\hbox{\hskip-3pt$\sim$}}
\raise1pt\hbox{$<$}}}     %less than approx. symbol
\def\gsim{\mathrel{\rlap{
\lower4pt\hbox{\hskip-3pt$\sim$}}
\raise1pt\hbox{$>$}}}     %greater than or approx. symbol
\def\be{\begin{eqnarray}}
\def\ee{\end{eqnarray}}

\begin{document}
%\today

\title{Event-by-event background  in estimates of the chiral magnetic effect}

\author{V. D. Toneev}
\affiliation{Joint Institute for Nuclear Research, 141980 Dubna, Russia}
\affiliation{Frankfurt Institute for Advanced Studies, 60438 Frankfurt am Main, Germany}

\author{V. P. Konchakovski}
\affiliation{Institute for Theoretical Physics, University of Giessen, 35392 Giessen, Germany}
\affiliation{Bogolyubov Institute for Theoretical Physics, 03680 Kiev, Ukraine}
\affiliation{Frankfurt Institute for Advanced Studies, 60438 Frankfurt, Germany}

\author{V. Voronyuk}
\affiliation{Joint Institute for Nuclear Research, 141980 Dubna, Russia}
\affiliation{Bogolyubov Institute for Theoretical Physics, 03680 Kiev, Ukraine}
\affiliation{Frankfurt Institute for Advanced Studies, 60438 Frankfurt, Germany}

\author{E. L. Bratkovskaya}
\affiliation{Institute for Theoretical Physics, University of Frankfurt, 60438 Frankfurt, Germany}
\affiliation{Frankfurt Institute for Advanced Studies, 60438 Frankfurt, Germany}

\author{W. Cassing}
\affiliation{Institute for Theoretical Physics, University of Giessen, 35392 Giessen, Germany}

\begin{abstract}
In terms of the parton hadron string dynamics (PHSD) approach ---
including the retarded electromagnetic field --- we investigate the
role of fluctuations of the correlation function in the azimuthal
angle $\psi$ of charged hadrons that is expected to be a sensitive
signal of local strong parity violation. For the early time we
consider fluctuations in the position of charged spectators resulting
in electromagnetic field fluctuations as well as in the position of
participant baryons defining the event plane. For partonic and
hadronic phases in intermediate stages of the interaction we study the
possible formation of excited matter in electric charge dipole and
quadrupole form as generated by fluctuations. The role of the
transverse momentum and local charge conservation laws in the observed
azimuthal asymmetry is investigated, too. All these above-mentioned
effects are incorporated in our analysis based on event-by-event PHSD
calculations. Furthermore, the azimuthal angular correlations from
Au+Au collisions observed in the recent STAR measurements within the
Relativistic Heavy Ion Collider (RHIC) Beam Energy Scan (BES) program
are studied. It is shown that the STAR correlation data at the
collision energies of $\sqrt{s_{NN}}=$ 7.7 and 11.5 GeV can be
reasonably reproduced within the PHSD. At higher energies the model
fails to describe the $\psi$ correlation data resulting in an
overestimation of the partonic scalar field involved. We conclude that
an additional transverse anisotropy fluctuating source is needed which
with a comparable strength acts on both in- and out-of-plane
components.
\end{abstract}

\maketitle

\section{Introduction}

A fundamental property of the non-Abelian gauge theory is the
existence of nontrivial topological configurations in the QCD
vacuum. Spontaneous transitions between topologically different states
occur with a change of the topological quantum number characterizing
these states and induce anomalous processes like local violation of
the ${\cal P}$ and ${\cal CP}$ symmetry. The interplay of topological
configurations with (chiral) quarks shows the local imbalance of
chirality. Such a chiral asymmetry when coupled to a strong magnetic
field induces a current of electric charge along the direction of the
magnetic field which leads to a separation of oppositely charged
particles with respect to the reaction plane
\cite{KMcLW07,Kh09,FKW08}.

This strong magnetic field can convert topological charge fluctuations
in the QCD vacuum into a global electric charge separation with
respect to the reaction plane.  Thus, as argued in
Refs.~\cite{KZ07,KMcLW07,FKW08,KW09}, the topological effects in QCD
might be observed in heavy-ion collisions directly in the presence of
very intense external electromagnetic fields due to the ``chiral
magnetic effect'' (CME) as a manifestation of spontaneous violation of
the ${\cal CP}$ symmetry. Indeed, it was shown that electromagnetic
fields of the required strength can be created in relativistic
heavy-ion collisions~\cite{KMcLW07,SIT09,EM_HSD} by the charged
spectators in peripheral collisions.

The first experimental evidence for the CME, identified via the charge
asymmetry, was obtained by the STAR Collaboration at the Relativistic
Heavy Ion Collider (RHIC) at $\sqrt{s_{NN}}=$ 200 and 62
GeV~\cite{Vol09,STAR-CME,STAR-CME2} and confirmed qualitatively by the
PHENIX Collaboration~\cite{PHENIX}.  Recently, these measurements were
extended, from one side, below the nominal RHIC energy down to
$\sqrt{s_{NN}}=$ 7.7 GeV within the RHIC Beam Energy Scan (BES)
program~\cite{BES11} and, from the other side, preliminary results for
the maximal available energy $\sqrt{s_{NN}}=$ 2.76 TeV were announced
from the Large Hadron Collider (LHC)~\cite{Ch11,LHC_CS12}.  Though at
first sight, some features of these data appear to be consistent with
an expectation from the local parity violation phenomenon, the
interpretation of the observed effect is still under intense
discussion~\cite{Wa09,BKL09,Pr10-1,Pr10,Pr09,AMM10,LKB10,BKL10,Lo11,Wa12}.

The fluctuation nature of the CME will give rise to a vanishing
expectation value of a ${\cal P}$-odd observable and due to that, as
proposed by Voloshin~\cite{Vol04}, the azimuthal angle two-particle
correlator related to charge asymmetry with respect to the reaction
plane is measured in
experiments~\cite{Vol09,STAR-CME,STAR-CME2,BES11,Ch11,LHC_CS12}. Accompanying
these experiments hadronic estimates of the dynamical background in
these experimental papers including only statistical (hadronic)
fluctuations do not involve the electromagnetic field at all. The
electromagnetic field --- created in heavy-ion collisions --- was
calculated in different dynamical approaches in
Refs.~\cite{KMcLW07,SIT09,EM_HSD,BES-HSD,TV10,OL11}. In two of
them~\cite{EM_HSD,BES-HSD} calculations were carried out in comparison
with the CME observable. However, in all these studies only the mean
electromagnetic field was presented, being averaged over the whole
ensemble of colliding nuclei.

As noted in Ref.~\cite{BS11} event-by-event fluctuations of the
electromagnetic field in off-central heavy-ion collisions can reach
rather high values comparable with the average values. The presence of
large fluctuations was then confirmed in a more elaborated model in
Ref.~\cite{DH12}. In this study --- based on the parton-hadron-string
dynamics (PHSD) kinetic approach --- we analyze event-by-event
fluctuations in the electromagnetic fields as well as in transverse
momentum, multiplicity and conserved quantities and the influence of
these fluctuations on physics observables relevant to measurements of
the CME.

The paper is organized as follows. After a short recapitulation of the
PHSD approach in Sec.~\ref{Sec:PHSD} we sequentially
(Sec.~\ref{Sec:Sources}) consider the manifestation of the initial
geometry fluctuations in spectator protons and participant nucleons as
well as in the charged quasiparticle geometry at some later
stage. These effects are relevant for fluctuations in the
electromagnetic fields, the event plane orientation and the possible
formation of a fluctuating electric charge dipole/quadrupole transient
subsystem, respectively. Conservation of the transverse momentum and
local charge is analyzed as an alternative explanation of the observed
azimuthal asymmetry. In Sec.~\ref{Sec:Observable} we discuss the role
and importance of these effects in the azimuthal angle correlations
and their dependence on collision energy. Our conclusions are
summarized in Sec.~\ref{Sec:Summary}.

\section{Reminder of the PHSD approach}
\label{Sec:PHSD}

Here we analyze the dynamics of partons, hadrons and strings in
relativistic nucleus-nucleus collisions within the parton hadron
string dynamics approach~\cite{PHSD}. In this transport approach the
partonic dynamics is based on Kadanoff-Baym equations for Green
functions with self-energies from the dynamical quasiparticle model
(DQPM)~\cite{Cassing06,Cassing07} which describes QCD properties in
terms of `resummed' single-particle Green functions. In
Ref.~\cite{BCKL11}, the actual three DQPM parameters for the
temperature-dependent effective coupling were fitted to the recent
lattice QCD results of Ref.~\cite{aoki10}.  The latter leads to a
critical temperature $T_c \approx$ 160~MeV which corresponds to a
critical energy density of $\epsilon_c \approx$ 0.5~GeV/fm$^3$. In
PHSD the parton spectral functions $\rho_j$ ($j=q, {\bar q}, g$) are
no longer $\delta-$functions in the invariant mass squared as in
conventional cascade or transport models but depend on the parton mass
and width parameters which were fixed by fitting the lattice QCD
results from Ref.~\cite{aoki10}. We recall that the DQPM allows one to
extract a potential energy density $V_p$ from the space-like part of
the energy-momentum tensor as a function of the scalar parton density
$\rho_s$.  Derivatives of $V_p$ with respect to $\rho_s$ then define a
scalar mean-field potential $U_s(\rho_s)$ which enters into the
equation of motion for the dynamic partonic quasiparticles. Thus, one
should avoid large local fluctuations in the potential $V_p$ which
indeed is solved in the parallel ensemble method by averaging the
mean-field over many events. In the present study we modify the
default PHSD approach by evaluating the electromagnetic fields for
each event without averaging the charge currents over many (parallel)
events.

The transition from partonic to hadronic degrees of freedom (d.o.f.)
(and vice versa) is described by covariant transition rates for the
fusion of quark-antiquark pairs or three quarks (antiquarks),
respectively, obeying flavor current-conservation, color neutrality as
well as energy-momentum conservation~\cite{PHSD,BCKL11}. Since close
to the phase transition the dynamical quarks and antiquarks become
very massive, the formed resonant `prehadronic' color-dipole states
($q\bar{q}$ or $qqq$) are of high invariant mass, too, and
sequentially decay to the ground-state meson and baryon octets
increasing the total entropy.

On the hadronic side PHSD includes explicitly the baryon octet and
decouplet, the $0^-$- and $1^-$-meson nonets as well as selected
higher resonances as in the hadron string dynamics (HSD)
approach~\cite{Ehehalt,HSD}.  Note that PHSD and HSD merge at low
energy density, in particular below the critical energy density
$\epsilon_c\approx$ 0.5~GeV/fm$^{3}$.

The PHSD approach has been applied to nucleus-nucleus collisions from
$\sqrt{s_{NN}}\sim$ 5 to 200~GeV in Refs.~\cite{PHSD,BCKL11} in order
to explore the space-time regions of `partonic matter'.  It was found
that even central collisions at the top-SPS energy of $\sqrt{s_{NN}}=$
17.3~GeV show a large fraction of nonpartonic, {\it i.e.}, hadronic or
string-like matter, which can be viewed as a hadronic corona. This
finding implies that neither hadronic nor only partonic `models' can
be employed to extract physical conclusions in comparing model results
with data.  All these previous findings provide promising perspectives
to use PHSD in the whole range from about $\sqrt{s_{NN}}=$ 5 to
200~GeV for a systematic study of azimuthal asymmetries of hadrons
produced in relativistic nucleus-nucleus collisions. This expectation
has been realized, in particular, in the successful description of
various flow harmonics in the transient energy
range~\cite{KBCTV11,KBCTV12}.

\begin{figure}[t]
\includegraphics[width=5.50truecm,clip=] {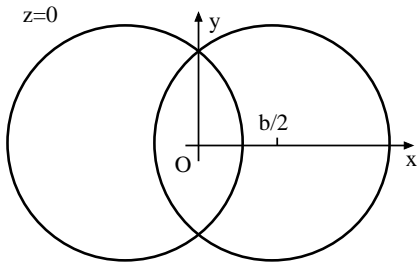}
\caption{The transverse plane of a noncentral heavy-ion collision.
  The impact parameter of the collision is denoted by $b$. The origin
  $O$ (corresponding to $x=0$) is referred to as the central point of
  the maximum overlap. }
 \label{tr-pl}
\end{figure}

The collision geometry for a peripheral collision is displayed in
Fig.~\ref{tr-pl} in the transverse $(x-y)$ plane. The reaction plane
is defined as the $(z-x)$ plane. The overlapping strongly interacting
region (participants) has an ``almond''-like shape. The nuclear region
outside this ``almond'' corresponds to spectator matter which is the
dominant source of the electromagnetic field at the very beginning of
the nuclear collision. Note that in the PHSD approach the particles
are subdivided into target and projectile spectators and participants
not geometrically but dynamically: spectators are nucleons which
suffered yet no hard collision.

As in Refs.~\cite{SIT09,EM_HSD,BS11,DH12} the electric and magnetic
fields at the relative position $\mathbf{R}_n={\bf r}-{\bf r}_n$ are
calculated according to the retarded $(t_n=t-|{\bf r}-{\bf r}_n|$)
Li\'enard-Wiechert equations for a charge moving with velocity
$\mathbf{v}$:
\be  \label{LWeq1}
 e\,\mathbf{E}(\mathbf{r}, t) &=&   \alpha \sum_n Z_n
 \frac{[{\bf R}_n-R_n{\bf v}_n]}{( R_n -{\bf R}_n\cdot{\bf v})^3} (1-v^2)~,
 \\\label{LWeq2}
 e\,\mathbf{B}(\mathbf{r}, t) &=& \alpha \sum_n Z_n
 \frac{{\bf v}\times{\bf R}_n}{( R_n -{\bf R}_n\cdot{\bf v})^3} (1-v^2)~,
\ee
where the summation runs over all charged quasiparticles in the
system, both spectators and participants, $Z_n$ is the charge of the
particle and $\alpha=e^2/4\pi=1/137$ is the electromagnetic
constant. By including explicitly the participants --- created during
the heavy-ion reaction and being propagated in time also under the
influence of the retarded electromagnetic fields --- we also consider
the back reaction of the particles on the retarded
fields. Equations~(\ref{LWeq1}), (\ref{LWeq2}) have singularities for
$R_n=0$ and in the calculations we regularize them by the condition
$R_n>$ 0.3 fm.

However, if the produced matter, after the short early-stage
evolution, is in the QGP phase, the electric conductivity is not
negligible. Strictly speaking our estimates of the magnetic and
electric fields in Eqs.~(\ref{LWeq1}), (\ref{LWeq2}) are strictly
valid only at the early stage of the collision. At later stages we
have neglected the collective electromagnetic response of the matter
produced in the collision by assuming that the produced matter is
ideally electrically insulating. Here, the magnetic response from the
created medium is expected to become increasingly important
\cite{LL84} and in principle may substantially influence the time
evolution of the electromagnetic fields in the QGP. In particular, a
non-trivial electromagnetic response --- as studied within generalized
Maxwell equations including the permeability and permittivity of the
QGP --- can lead to a slowdown of the decrease of the magnetic field
at later times of 2-4 fm/c \cite{DH12,Tu10}.  It is of interest to
recall that for a peripheral Au+Au collision at $\sqrt{s_{NN}}=$ 200
GeV our kinetic model with the retarded electromagnetic field predicts
a flattening of the strong time dependence for the magnetic field at
$t\approx$1 fm/c (see Fig.~4 in Ref.~\cite{EM_HSD}). It is also
noteworthy that the magnetic field strength at this time is by three
orders of magnitude lower than the maximal field strength. Therefore
it is not likely that there will be a noticeable influence of the
effect discussed above on observables for later times. Furthermore, we
mention that according to Faraday's law a strongly decreasing magnetic
field induces an electric field circulating around the direction of
the magnetic field.  In turn this electric field generates an electric
current that produces a magnetic field pointing in the positive z
direction according to the Lenz rule~\cite{Tu11}. All these nontrivial
responses of charged matter to intense electromagnetic fields are of
great interest and more elaborated studies are required; however, this
is beyond the aim of this present paper.

\section{Sources of background fluctuations}
\label{Sec:Sources}

\subsection{Fluctuations in the proton spectator positions}
\label{glasma}

\begin{figure*}[thb]
\begin{center}
\includegraphics[width=0.4\textwidth,clip]{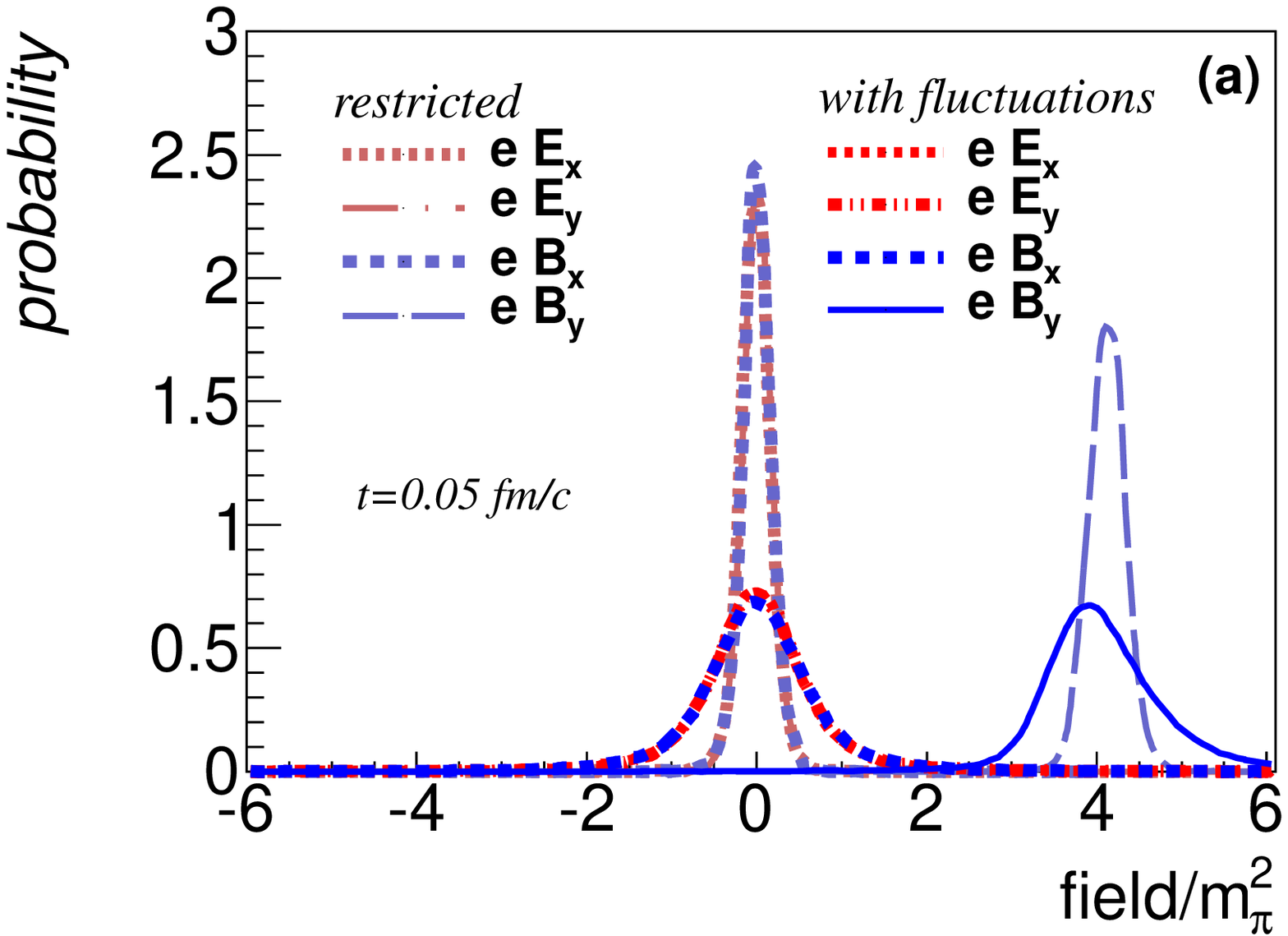}
\hspace*{8mm}\includegraphics[width=0.4\textwidth,clip]{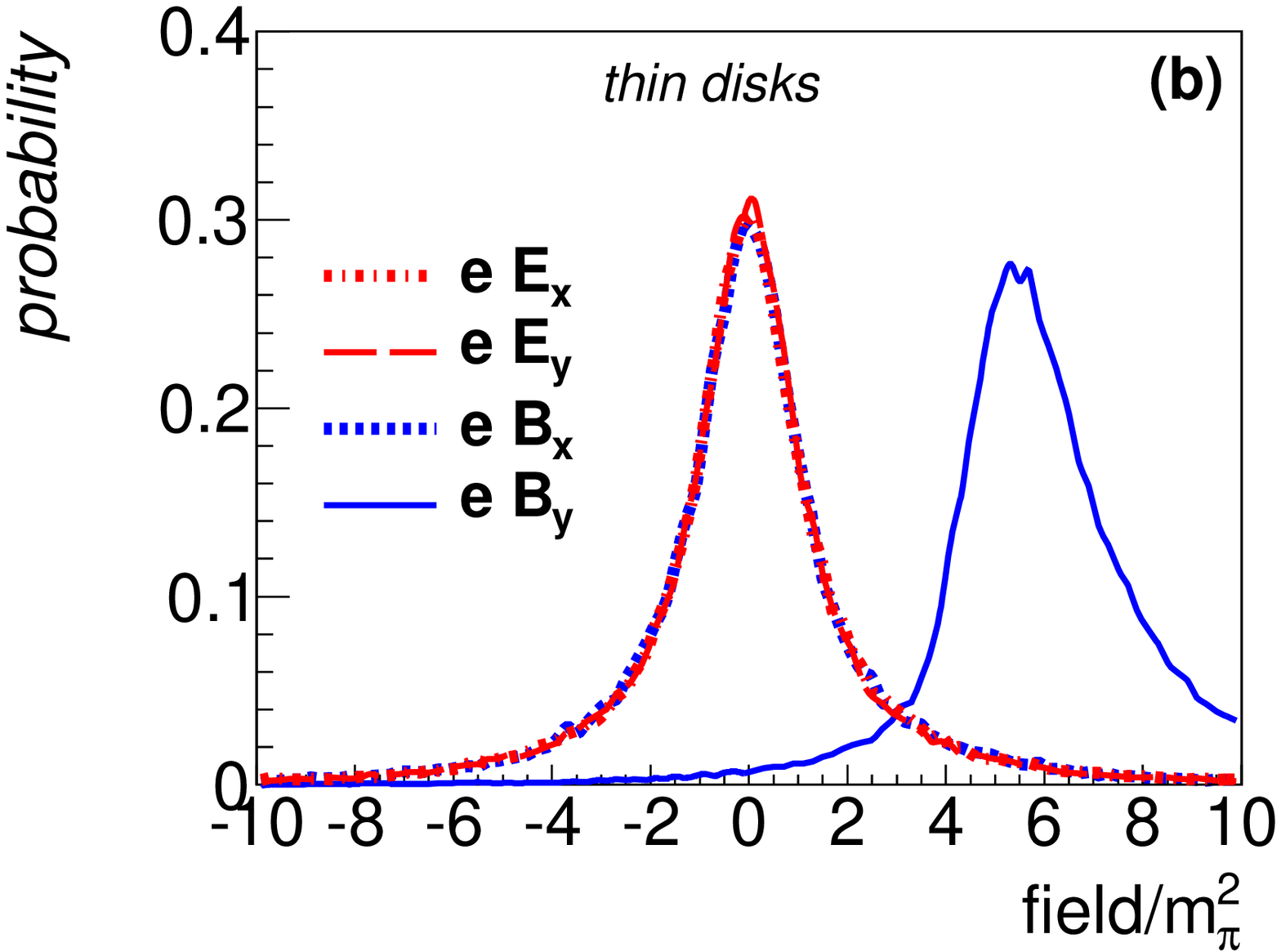}
\end{center}
\caption{(Color online) Probability densities for electromagnetic
  fields at the central point of off-central Au+Au collisions for
  $\sqrt{s_{NN}}=$ 200 GeV at the time of maximal overlap as emerging
  from event-by-event calculations. The results are given for the PHSD
  model (a) and for the schematic model with infinitely thin nuclei as
  in Ref.~\cite{BS11} (b).}
\label{Pfield}
\end{figure*}

Let us consider geometrical fluctuations in the electromagnetic field
taking into account fluctuations in the position of {\it spectator}
protons. The retarded electric and magnetic field evaluated according
to Eqs.~(\ref{LWeq1}),(\ref{LWeq2}) are presented in Fig.~\ref{Pfield}
for off-central Au+Au collisions at the collision energy of
$\sqrt{s_{NN}}=$ 200 GeV. The PHSD results (including contributions of
all quasiparticles) are given for the time of the maximal overlap of
the compressed colliding nuclei which corresponds to $t\simeq$ 0.05
fm/c. As noted above the main contribution is coming from spectator
protons. In peripheral collisions the average magnetic component
orthogonal to the reaction plane $<B_y>$ is dominant. The
dimensionless field magnitude $e<B_y>/ {m_\pi}^2\simeq$ 5 and its
dispersion are in a reasonable agreement (discrepancy is less than
10\%) with recent calculations results within the partonic HIJING
model~\cite{DH12}.  The difference in the calculated electromagnetic
field between HIJING and our PHSD approach is due to different
regularization procedures used for Eqs.~(\ref{LWeq1}),
(\ref{LWeq2}). In \cite{DH12} all field contributions resulting in a
numerical overflow were taken away while we used a constraint on the
closest distance $R_n>0.3$ fm. The agreement between these two models
demonstrates a very week sensitivity of the results for reasonable
values $R_n$. Our present results are consistent also with earlier
calculations within the hadronic dynamics of the ultrarelativistic
quantum-molecular dynamics (UrQMD)~\cite{SIT09} and the hadron string
dynamics (HSD)~\cite{EM_HSD} models.

If one looks at the field variance [Fig.~\ref{Pfield}(a)] the full
width of the $E_y, E_x, B_x$ distributions is about $\sigma \sim 2
/m_\pi^2$ for all transverse field components being consistent with
Ref.~\cite{DH12}.  Here, additional results are plotted also for the
restricted case when the electromagnetic field is averaged over all
events in the parallel ensemble as explained in the previous
section. This procedure has been used before in Ref.~\cite{EM_HSD}. As
seen from Fig.~\ref{Pfield}(a) this leads to a suppression of the
variance for all field distributions by a factor of about 3.

In Fig.~\ref{Pfield}(b) we mimic results of the schematic model in
Ref.~\cite{BS11} considering a nuclear colliding system at the time of
the maximal overlap as an infinitely thin disk. This was simulated
numerically by an artificial shift of the position of the longitudinal
components of all protons at this moment to the plane $z=0$. As is
seen in Fig.~\ref{Pfield}(b) all field distributions indeed increase
in width by a factor of about two. A direct comparison of our results
to those of Ref.~\cite{BS11} gives a factor of three or even
more. This finding completely coincides with the results of
Ref.~\cite{DH12} as to both the value of the width and its origin.

\begin{figure}[thb]
\includegraphics[width=0.37\textwidth,clip] {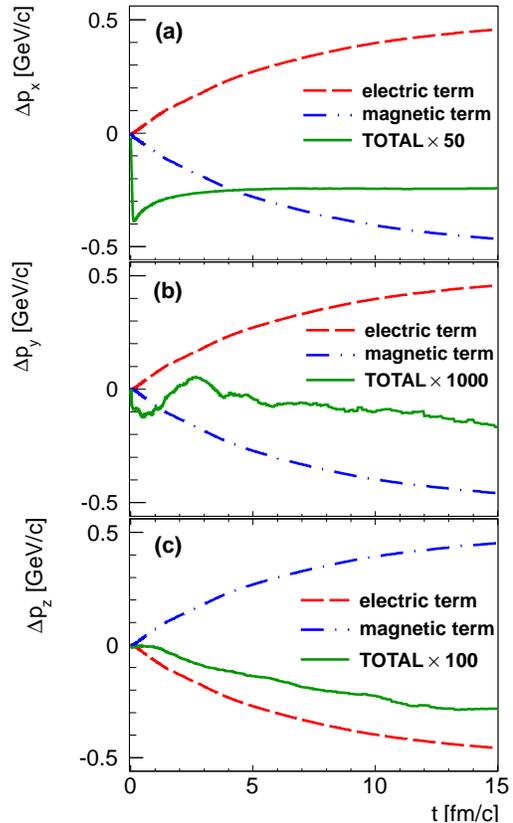}
\caption{(Color online) Time dependence of the momentum increment of
  forward moving ($p_z>0$) partons due to the electromagnetic field
  created in Au+Au ($\sqrt{s_{NN}}=$ 200 GeV) collisions with the
  impact parameter $b=$ 10 fm.}
\label{Dp-150}
\end{figure}

\begin{figure*}[thb]
\includegraphics[height=6.0truecm] {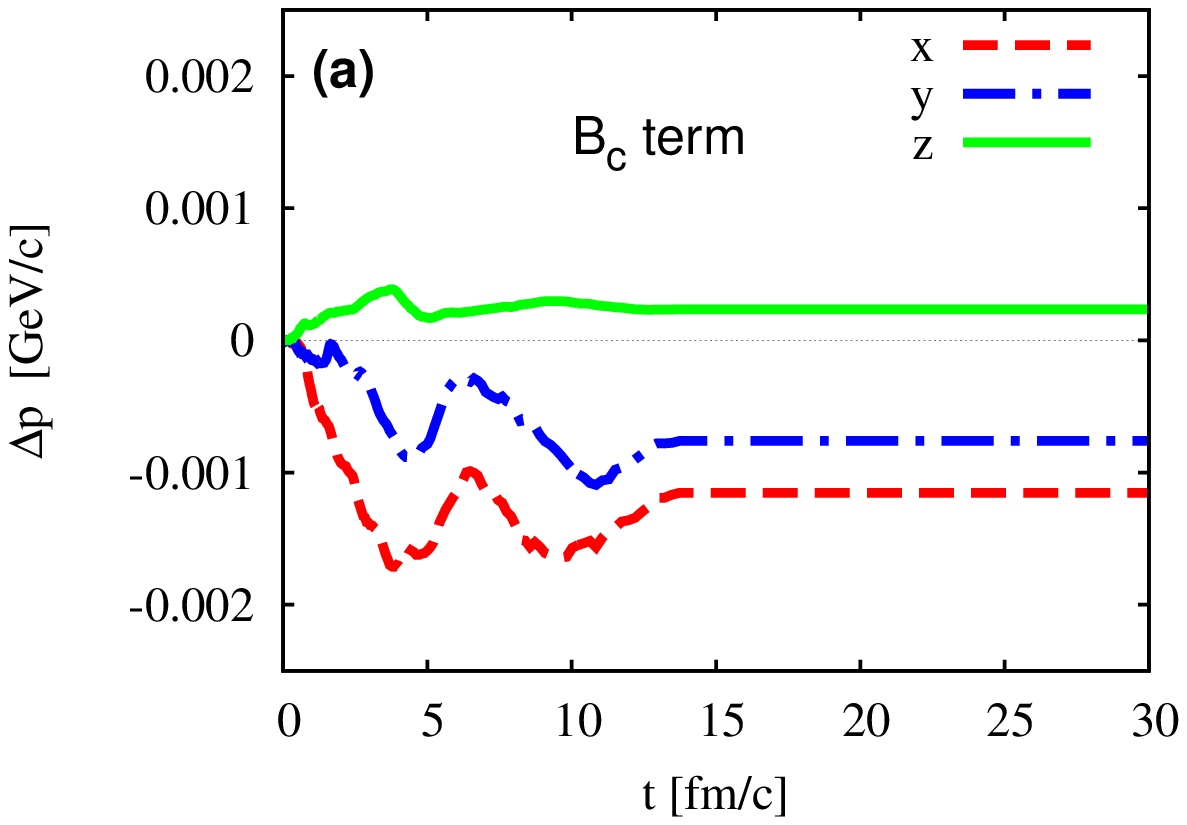}
\includegraphics[height=6.0truecm] {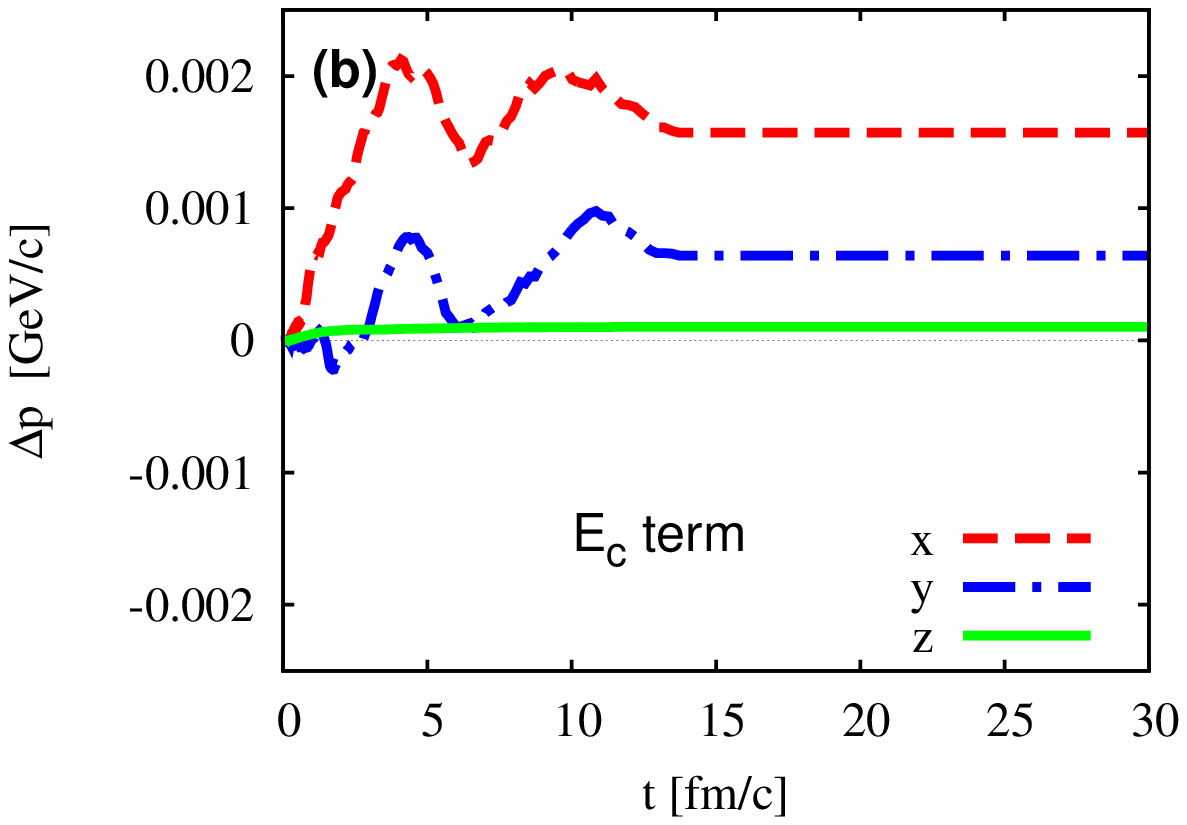}
\includegraphics[height=6.0truecm] {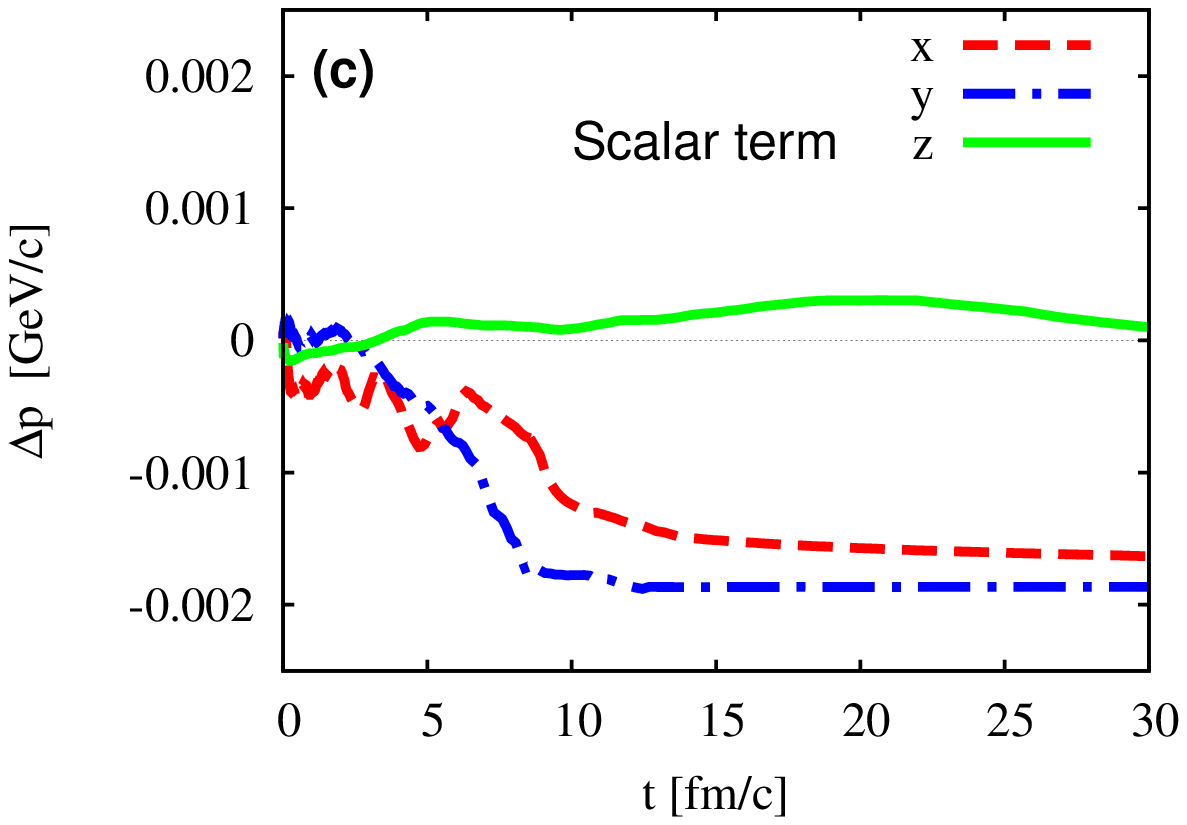}
\includegraphics[height=6.0truecm] {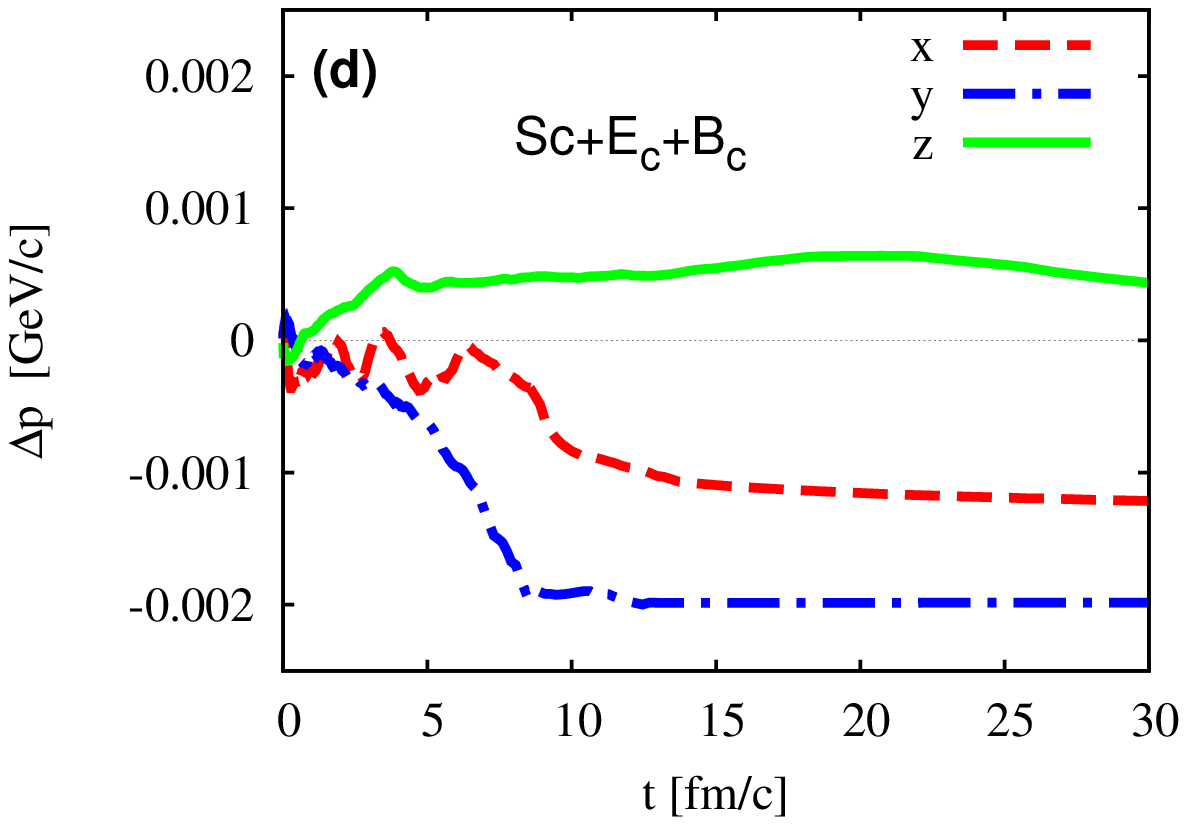}
\caption{(Color online) Time dependence of the momentum increment from
  `electric' $E_c$ and `magnetic' $B_c$ (a), (b) partonic field
  components, the scalar field and total momentum increment (c), (d)
  for forward moving ($p_z>0$) positively charged quarks and $p>0$ as
  a function of time $t$. The system is Au+Au (at $\sqrt{s_{NN}}=$ 200
  GeV) for the impact parameter $b=$ 10 fm.}
\label{Dp-cromo-comp}
\end{figure*}

The estimated strength of the electromagnetic fields provides no
information about their action on the quasiparticle transport. Let us
look at the early time dynamics in more detail and introduce a
momentum increment $\Delta {\bf p}$ as a sum of the mean particular
increases of the quasiparticle momentum $d{\bf p}$ due to the action
of the electric and magnetic forces,
\be \label{force} 
{\bf F}_{em}=e{\bf E}+(e/c) \,{\bf v}\times {\bf B},
\ee
during the short time interval at the expense of the given source,
\be \label{dt}
 \Delta {\bf p}(t)=\sum_{t_i}^t \langle d{\bf p}(t_i) \rangle .
\ee
Equation (\ref{dt}) is considered on an event-by-event basis and for
each event the mean momentum increase during a time-step $\langle
d{\bf p}(t_i) \rangle$ is calculated over all particles involved. In
Fig.~\ref{Dp-150} the average momentum change of forward moving quarks
$(p_z>0)$ is shown for three components of the electromagnetic force
at $\sqrt{s_{NN}}=$ 200 GeV. Note the different scales for the solid
lines in Fig.~\ref{Dp-150} that give the net momentum change at this
energy. It is a remarkable fact that the transverse electric and
magnetic components compensate each other almost completely.

Two remarks are in order: First, due to the linearity of the
electromagnetic force (\ref{force}) with respect to the electric and
magnetic field, one should not expect a difference in quark transport
calculations with and without taking into account electromagnetic
field fluctuations. This was demonstrated for quasiparticles earlier
in terms of the HSD model~\cite{BES-HSD}.  Second, if transverse
fluctuations are characterized by the average strength of the fields,
$\langle|E_{x,y}|\rangle$ and $\langle|B_{x,y}|\rangle$, certain
equalities between components like $\langle|E_{x}|\rangle\approx
\langle|E_{y}|\rangle\approx \langle|B_{x}|\rangle$ --- as numerically
obtained in Ref.~\cite{BS11} and confirmed in Ref.~\cite{DH12} ---
imply that similar equalities should hold for the
fluctuations. Indeed, similar relations follow from our PHSD
calculations, see Fig.~\ref{Pfield}(a) where the increment functions
for appropriate field components practically coincide.  We emphasize
again that the PHSD transverse field components are not only of
comparable strength but their action on the quarks [see
Eq. (\ref{force})] approximately compensate each other.  One should
note that this is the compensation effect rather than the short
lifetime of the electromagnetic interaction which leads to a very weak
sensitivity of observables as has been demonstrated recently in terms
of the hadronic HSD transport model in Ref.~\cite{BES-HSD}. For a
quasiparticle moving along the trajectory $x=x(t)$, this compensation
in a simplified 1D case can be illustrated by a short calculation as
\be
  e E=-e\frac{\partial A}{\partial t}
  \sim -e\frac{\partial A}{\partial x} \frac{dx}{dt} \sim -eBv~, \label{comp}
\ee
{\it i.e.}, the action of the electric and magnetic transverse
components is roughly equal and directed oppositely.

\begin{figure*}[bht]
\includegraphics[height=12.cm,width=0.30\textwidth, clip]{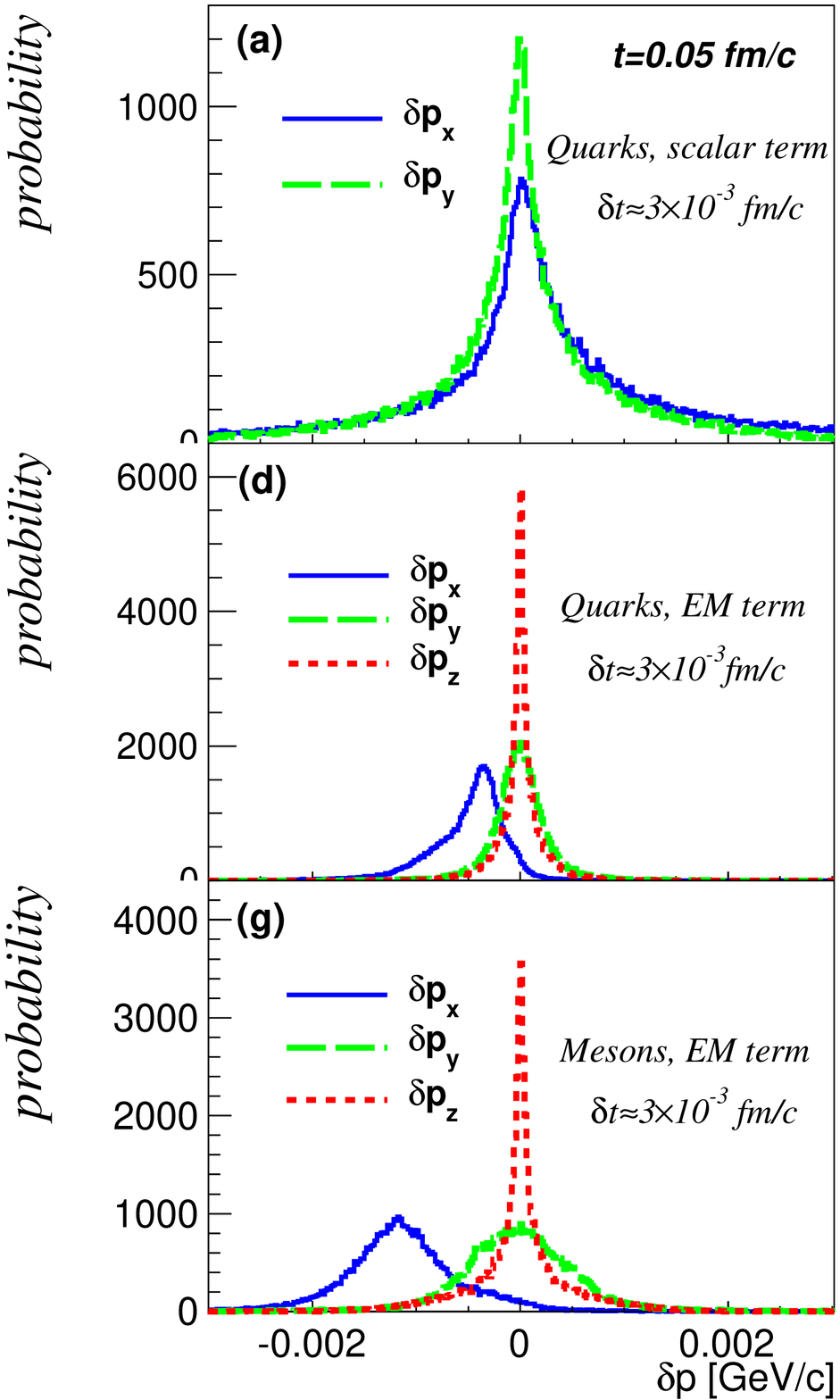}
\includegraphics[height=12.cm,width=0.30\textwidth, clip]{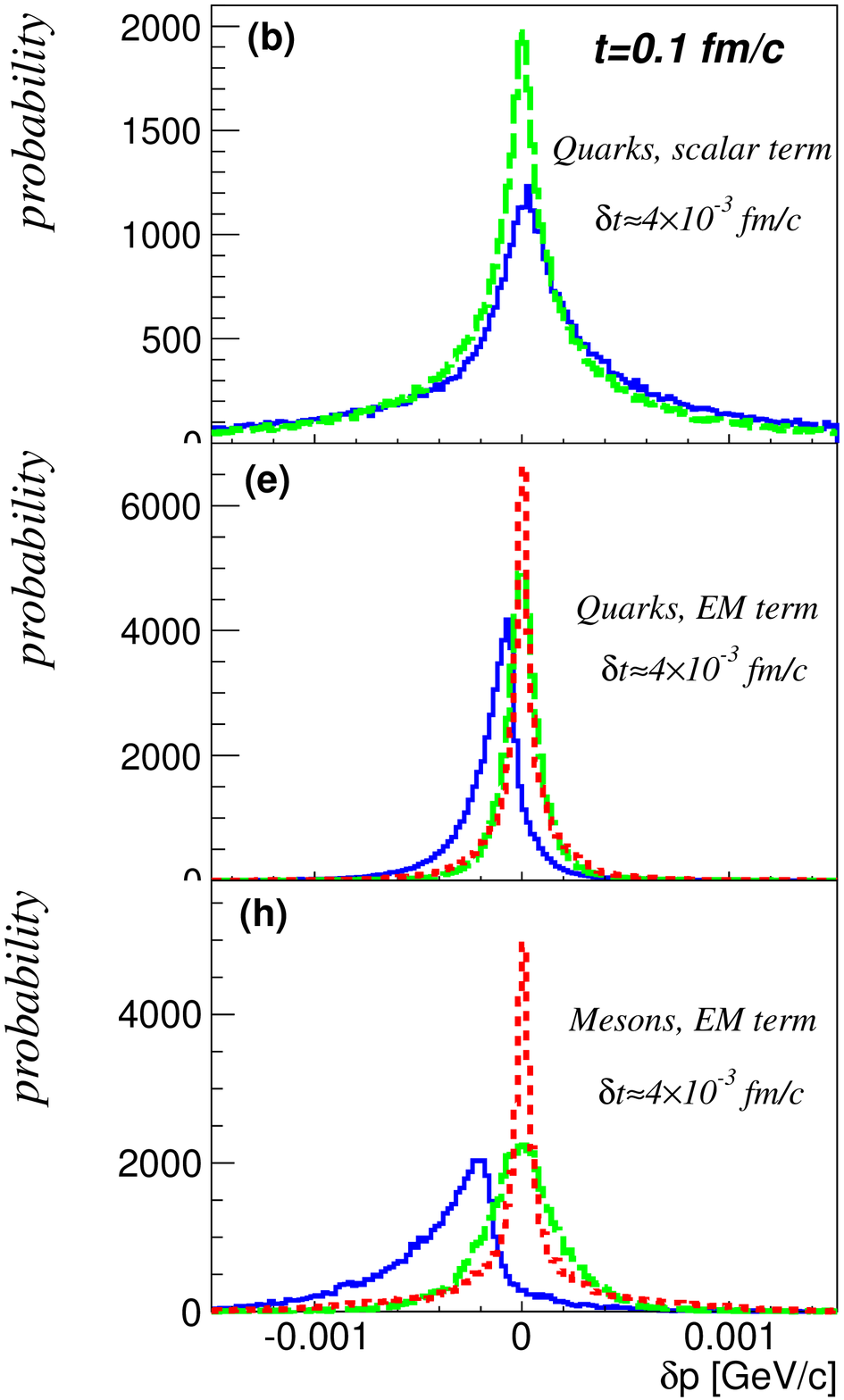}
\includegraphics[height=12.cm,width=0.30\textwidth, clip]{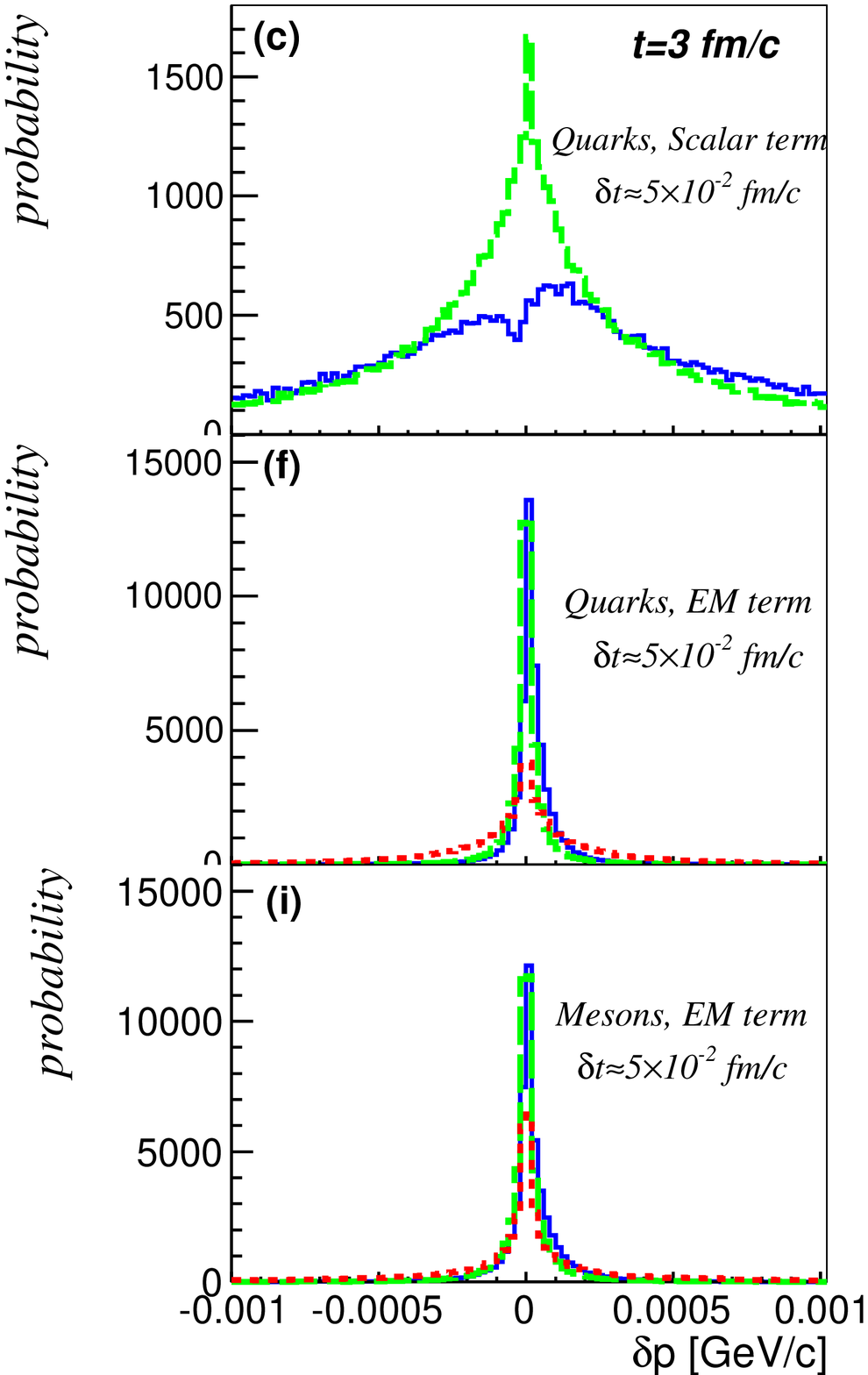}
\caption{(Color online) Probability distribution of the momentum
  increment $\delta {\bf p}={\bf p}-\langle {\bf p}\rangle$ during the
  time step $\delta t$ for forward moving positively charged
  quasiparticles at times $t=$ 0.05, 0.1, and 3 fm/c. The distribution
  emerging from the quark scalar potential is shown in the upper
  panels (a),(b),(c) whereas the distribution stemming from the EM
  field of quarks and mesons are displayed in the middle (d),(e),(f)
  and bottom (g),(h),(i) panels, respectively. The calculations have
  been performed for off-central Au+Au collisions at $\sqrt{s_{NN}}=$
  200 GeV and impact parameter $b=$ 10~fm in the PHSD model.}
\label{dp1}
\end{figure*}

The important advantage of the PHSD approach relative to hadron-string
models is the inclusion of partonic degrees of freedom. In particular,
the involved partonic fields (of scalar and vector type) showed up to
be essential to describe the elliptic flow excitation function from
lower SPS to top RHIC energies~\cite{KBCTV11,KBCTV12} and to be a key
quantity in analyzing the CME.

The evolution of momentum increments for partonic forces is presented
in Fig.~\ref{Dp-cromo-comp} for off-central Au+Au collisions at
$\sqrt{s_{NN}}=$ 200 GeV. It is seen that (marked by the subscript
$``c"$) the transverse `electric' $E_c$ and `magnetic' field $B_c$ of
the partonic field components [Fig.~\ref{Dp-cromo-comp}(a),(b)] almost
compensate each other. The $z$ component is practically vanishing and
for $t\gsim 8$ fm/c all quark increments stay roughly constant, {\it
  i.e.}, the quark phase ends here. The final action of the partonic
forces is defined by the sum of the forces (d) which is dominated by
the scalar one.

Apart from the average forces (momentum increments) the fluctuations
of the forces are of further interest. As seen from Fig.~\ref{dp1} the
distribution in the quark momentum deviation $\delta {\bf p}={\bf
  p}-\langle {\bf p}\rangle$ in case of scalar forces is well
collimated with respect to the average trajectory $\langle {\bf
  p}\rangle$ presented in Fig.~\ref{Dp-cromo-comp} but its width
increases by about a factor of three when proceeding from $t=$ 0.05 to
3.0 fm/c [Fig.~\ref{dp1}(a)-(c)]. This spread is slightly larger in
the $x$ component since the derivatives of the scalar mean-field are
higher in the $x$ than in the $y$ direction. The influence of the
electromagnetic force on quarks and charged pions is visible more
clearly [Fig.~\ref{dp1}(d)-(i)] in the early time corresponding to the
maximal overlap of the colliding nuclei ($t=$ 0.05 fm/c) when the
created electromagnetic field is maximal. Here, the $\langle \delta
p_x\rangle$ component is shifted for quarks (d)-(f) and
even more for mesons (g)-(i). This shift decreases in
time and disappears for $t=$ 3 fm/c; at this time the deviation
distributions for all three components of the electromagnetic force
are close to a $\delta$ function.

\begin{figure*}[thb]
\includegraphics[height=9.0truecm] {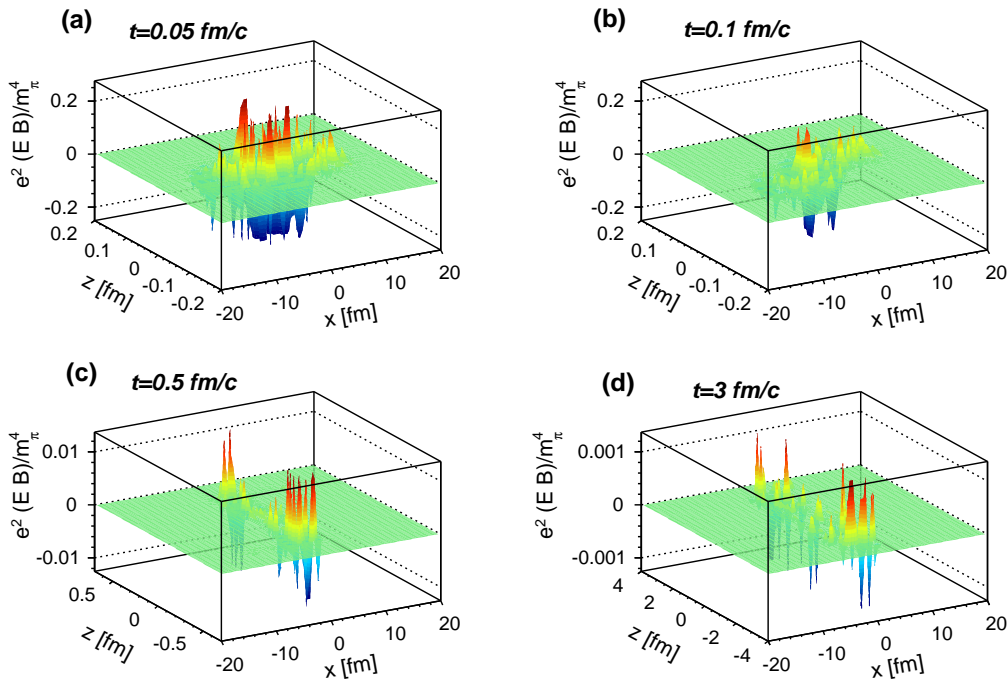}
\caption{(Color online) Space-time evolution of the scalar product of
  electric and magnetic fields $({\bf E}\cdot {\bf B})$ for Au+Au
  reactions at impact parameter $b=$ 10 fm and $\sqrt{s_{NN}}=$ 200
  GeV. Note the different scales along the $z$ axis.}
\label{EtoB}
\end{figure*}

Some general considerations on parity violation in heavy-ion reactions
are in order here: Since the magnetic field is odd under time reversal
(or equivalently, under the combined charge conjugation and parity
${\cal CP}$ transformation), the time reversal symmetry of a quantum
system is broken in the presence of an external magnetic field. A
magnetic field ${\bf B}$ can also combine with an electric field ${\bf
  E}$ to form the Lorentz invariant $({\bf E}\cdot {\bf B})$ which
changes the sign under a parity transformation. In the normal QCD
vacuum with its spontaneously broken chiral symmetry the leading
interaction involves the invariant $({\bf E}\cdot {\bf B})$ which
enters {\it e.g.}, into the matrix element that mediates the
two-photon decay of the neutral pseudoscalar mesons. In the deconfined
chirally symmetric phase of QCD, the leading interaction term is
proportional to $\alpha\alpha_s({\bf E}\cdot {\bf B})({\bf E^a}\cdot
{\bf B^a})$, where ${\bf E^a}$ and $ {\bf B^a}$ denote the
chromoelectric and chromomagnetic fields, respectively, and $\alpha_s$
is the strong QCD coupling. Both interactions are closely related to
the electromagnetic axial anomaly, which in turn relates the
divergence of the isovector axial current to the pseudoscalar
invariant of the electromagnetic field (see Ref.~\cite{MS10}). The
evolution of the electromagnetic invariant ${\bf E}\cdot {\bf B}$ is
shown in Fig.~\ref{EtoB}. The case of Au+Au ($\sqrt{s_{NN}}=$ 200 GeV)
collisions at impact parameter $b=$ 10 fm is considered.  As seen from
Fig.~\ref{EtoB} the electromagnetic invariant $({\bf E}\cdot {\bf B})$
is non-zero only in the initial time $t\lsim$ 0.5 fm/c where the
$({\bf E}\cdot {\bf B})$ distribution is quite irregular and its
nonzero values correlate well with the location of the nuclear overlap
region.  For later times this electromagnetic invariant vanishes in
line with the electric field space-time distributions~\cite{EM_HSD}.
Note that the quantities plotted in Fig.~\ref{EtoB} are dimensionless
and the scaling factor $m_\pi^4 \ [\text{GeV}^4]$ is quite small.

One should note that in addition to the strong electromagnetic
fields~\cite{KMcLW07,SIT09} present in noncentral collisions, very
strong color electric ${\bf E}^a$ and color magnetic ${\bf H}^a$
fields are produced in the very beginning of these collisions as shown
in the non-Abelian field theory~\cite{Fu12}. These fields can be
characterized by a gluon saturation momentum $Q_s$ and the time $\sim
1/Q_s$. Both fields are parallel to each other and directed along the
$z$ axis. This leads to a nonzero topological charge $Q \sim ({\bf
  E}^a\cdot {\bf B}^a)\ne$0. Since gauge fields with $Q\ne$ 0 generate
chirality, they also can induce electromagnetic currents along a
magnetic field~\cite{FKW10} resulting in the CME.  Though a large
amount of topological charge might be produced through the mechanism
of sphaleron transitions, the primary mechanism for topological charge
$Q$ generation at the early stage is by fluctuations of color electric
and magnetic fields. The decay of these fields is essentially governed
by the non-Abelian dynamics of the glasma~\cite{Fu12,LMcL06} which
ultimately produces the QGP (close to equilibrium). Unfortunately,
this possible mechanism for the CME is beyond the potential of the
PHSD model used. We thus may speculate about but not prove this
mechanism.

\subsection{Fluctuations in the position of participant nucleons}

\begin{figure}[tb]
\includegraphics[width=\linewidth]{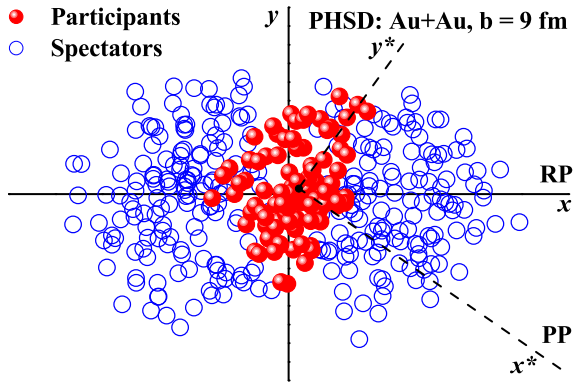}
\caption{(Color online) Projection of a single peripheral Au+Au
  (200~GeV) collision on the transverse plane. Spectator and
  participant nucleons are plotted by empty and filled circles,
  respectively. The reaction plane (RP) projection corresponds to $x$
  axis. Transverse axes of the participant plane (PP) are marked by
  stars ($x^\star,y^\star$).}
\label{C2}
\end{figure}

As noted above (see Fig.~\ref{tr-pl}), the interaction region after
averaging over many events has an almond-like shape; the averaged
spatial initial asymmetry of the participant matter is symmetric with
respect to the reaction plane.  Actual collision profiles, however,
are not smooth and the symmetry axis in an individual event is tilted
due to fluctuations (cf. Fig.~\ref{C2}). The geometry fluctuations in
the location of the {\it participant} nucleons lead to fluctuations of
the participant plane (PP) from one event to another, rendering larger
coordinate space eccentricities which due to pressure gradients are
translated into elliptic flow for the final state particles. Thus, the
system of the elliptical almond-like shape expands predominantly along
the minor axis.

Depending on the location of the participant nucleons in the colliding
nuclei at the time of the collision, the actual shape of the overlap
area may vary.  As is seen from Fig.~\ref{C2}, due to fluctuations the
overlap area in a single event can have, for example, a rotated
triangular rather than an almond shape. Note that an almond shape is
regained by averaging over many events for the same impact
parameter. However, in experiment the collective flows are measured
with respect to a third plane, the so-called {\it event} plane defined
by observable charged participants in momentum space through the
harmonic/multipole analysis. More precisely, the flow coefficients
$v_n$ are defined as the $n$th Fourier harmonic of the particle
momentum distribution with respect to the particular momentum event
plane $\Psi_n$,
\be
 \label{vn} \langle v_n\rangle = \langle
\cos[n(\psi-\Psi_n)]\rangle~,
\ee
where $\psi = \arctan(p_y/p_x)$ is the azimuthal angle of the particle
momentum ${\bf p}$ in the c.m.\ frame and angular brackets denote a
statistical average over many events.

\begin{figure}[tb]
\includegraphics[width=0.45\textwidth,clip]{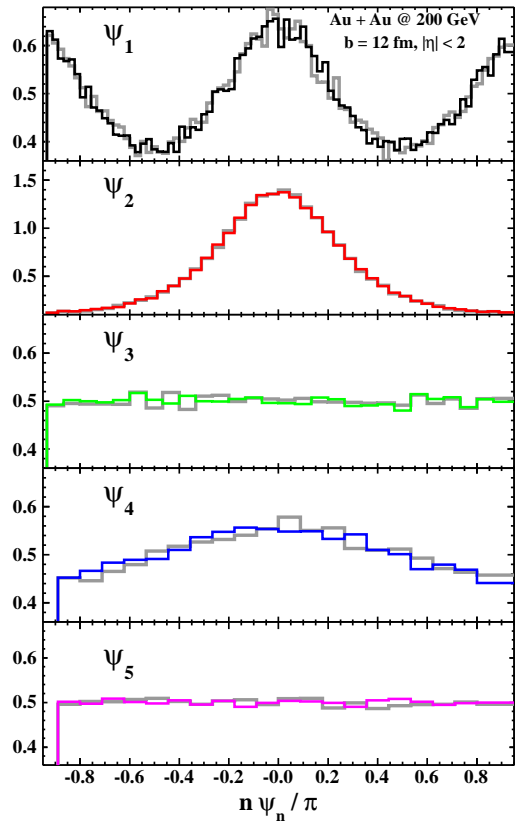}
\caption{(Color online) Distribution in the event plane angle for
  different harmonics $\Psi_n$ calculated with retarded magnetic and
  electric fields. Grey histograms show the results for respective
  calculations without fields.}
\label{Psi}
\end{figure}

One should note that all azimuthal correlations are not only due to
the collective flow.  The early-time two-particle spatial correlations
probe both the event geometry (fluctuating in individual events) and
genuine local pair correlations referred to as `nonflow'
correlations. The Fourier decomposition~(\ref{vn}) is not enough to
disentangle these two contributions. A possible solution of the
connection between flow fluctuations and initial state correlations is
given by the cumulant expansion method~\cite{BDO01} using two- and
four-particle correlation measurements of the harmonic flow
coefficients. However, this method is beyond the scope of the present
study.

The distributions in the event plane angle for different harmonics are
shown in Fig.~\ref{Psi} for the freeze-out case. All distributions are
symmetric with respect to the point $\Psi_n=0$ which corresponds to
the true reaction plane. As is seen, the event plane angle $\Psi_n$
determined from the $n$th harmonic is in the range $0\leq \Psi_n <
2\pi/n$ and fluctuations of several lowest order harmonics have
comparable magnitudes. Inside this region $\Psi_1$ has two maxima at
$\Psi_n=0$ and $\pi$ corresponding to forward-backward emission. The
even components $\Psi_2,\Psi_4$ have a rather prominent maximum for
$\Psi_n =0$ indicating the local nature of fluctuations, but the odd
harmonics $\Psi_3, \Psi_5$ are practically flat. This may be easily
understood since the odd moments of the spatial anisotropy purely
originate from fluctuations while the even ones are combined effects
of fluctuations and geometry. As a consequence, if one defines the
spatial anisotropy parameters with respect to the pre-determined
reaction plane, the event-averaged ones vanish for all odd moments but
not for the even moments.

The histograms in Fig.~\ref{Psi} are calculated from a sample of
$3\times 10^4$ events taking into account magnetic and electric field
fluctuations. Similar calculations without fields are shown in the
same figure by the grey histograms which are hard to distinguish from
the previous ones. In other words, there is no additional ``tilting''
effect by electromagnetic fields as expected in
Refs.~\cite{BS11,DH12}. This is due to the compensation of the
transverse electromagnetic components as explained above.

\subsection{Two-particle angular correlations}

An experimental signal of the local spontaneous parity violation is a
charged particle separation with respect to the reaction
plane~\cite{Vol05}. It is characterized by the two-body correlator in
the azimuthal angles,
\begin{eqnarray}
\label{cos}\gamma_{ij}&\equiv&\langle \cos (\psi_i+\psi_j-2\Psi_{RP})
\rangle \\ \nonumber &=&\langle \cos (\psi_i-\Psi_{RP})\ \cos (\psi_j -\Psi_{RP})\rangle
\\ \nonumber
&-&  \langle\sin (\psi_i-\Psi_{RP})\ \sin (\psi_j-\Psi_{RP})\rangle
\end{eqnarray}
where $\Psi_{RP}$ is the azimuthal angle of the reaction plane defined
by the beam axis and the line joining the centers of the colliding
nuclei and subscripts of $\gamma_{ij}$ represent the signs of electric
charges being positive or negative. The averaging in Eq. (\ref{cos})
is carried out over the whole event ensemble and $\cos$ and $\sin$
terms in Eq.~(\ref{cos}) correspond to out-of-plane and in-plane
projections of $\gamma_{ij}$.

As was proposed in Refs.~\cite{Pr10-1,Pr09} and more elaborated in
Ref.~\cite{BKL10}, a possible source of azimuthal correlations among
participants is the conservation of the transverse momentum which
might give rise to a contribution comparable with the measured CME.
Transverse momentum conservation (TMC) introduces back-to-back
correlations for particle pairs because they tend to balance each
other in transverse momentum space. A large multiplicity of particles
will dilute the effect of these two-particle correlation. Furthermore,
this correlation should be stronger in plane than out of plane due to
the presence of the elliptic flow. Nevertheless, the TMC provides a
background for the CME that should be properly quantified.

From quite general considerations --- making use of the central limit
theorem and describing particles thermodynamically --- one can derive
the following simple expression for the two-particle
correlator~\cite{BKL10}:
\be \label{TMC}
 \gamma_{ij} =
 \langle \cos(\psi_i+\psi_j)\rangle=-\frac{v_2 \ \langle p_t\rangle^2_{acc}}
  {N \ \langle p_t^2 \ \rangle_{full}}~,
\ee
where $N=N_++N_0+N_-$ is the total number of all produced particles
(in full phase space). At $\sqrt{s_{NN}}=$ 200 GeV it can be
approximated as $N \approx (3/2)\ N_{ch} \approx 21
\ N_{part}$~\cite{BKL10} where $N_{part}$ is calculated dynamically in
our model as well as the momentum-dependent factors for full phase
space and the ratio to the measured accepted phase space. It is of
interest to note that the proportionality of the CME to the elliptic
flow $v_2$ seen in Eq.~(\ref{TMC}) follows also from more elaborated
considerations. In particular, the chiral magnetic effect in the
hydrodynamic approach and in terms of a holographic gravity dual model
(see Ref.~\cite{GKK12}) predicts a linear dependence of the CME on
$v_2$ with more sophisticated coefficients which depend on the axial
anomaly coefficient and the axial chemical potential as well as on
dynamics of fluids through the particle density, baryon chemical
potential and pressure.

\begin{figure}[tb]
\begin{center}
\includegraphics[width=0.40\textwidth,clip]{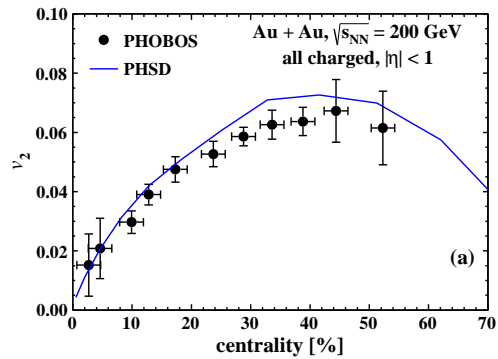}
\includegraphics[width=0.40\textwidth,clip]{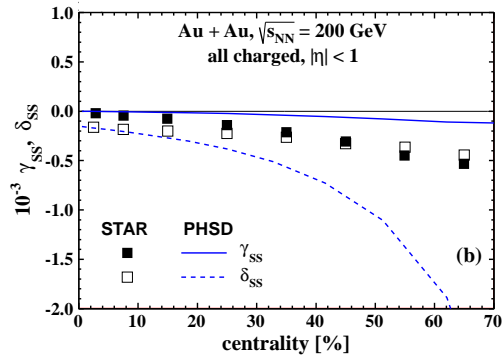}
\includegraphics[width=0.40\textwidth,clip]{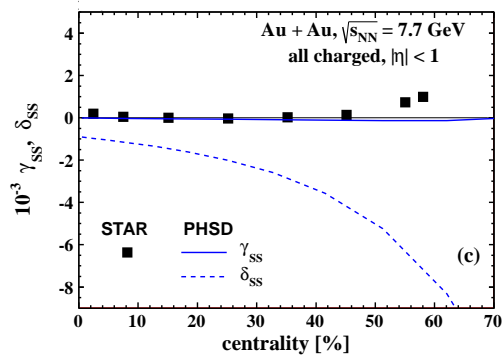}
\end{center}
\caption{(Color online) PHSD centrality dependence of the elliptic
  flow (a) and angular correlators $\gamma_{ss}$ and $\delta_{ss}$ of
  charged particles from Au+Au at $\sqrt{s_{NN}}=$ 200 (b) and 7 (c)
  GeV from the transverse momentum conservation according to
  Eqs.~(\ref{ss}) and (\ref{delta}).  The experimental data points for
  $ v_2$ and $\gamma_{ss},\ \delta_{ss}$ are from
  Refs.~\cite{Back:2004mh} and \cite{BES11}, respectively.}
\label{TMC-cos200}
\end{figure}

Experimentally, the same-sign correlator $\gamma_{ss}$ is defined as
the average of $\gamma_{++}$ and $\gamma_{--}$ by assuming that the
momentum balance is shared equally among the charges
 \be \label{ss}
 \gamma_{ss}=\frac{1}{2}(\gamma_{++}+\gamma_{--})=-\frac{v_2 \
 \langle p_t\rangle^2_{acc}}  {N \ \langle p_t^2 \
 \rangle_{full}}~,
 \ee
where the subscripts ``full" and ``acc" imply that average values should
be calculated in the ``full" phase space or in the proper ``acceptance
region", respectively. In practice, only a subset of particles is
measured. In this case some of the momentum balance stems from
unmeasured particles and one might expect $\gamma_{ss} \ll
v_2/N$~\cite{Pr09}. In the STAR experiment~\cite{Aggarwal:2010ya}
tracks were measured for the central two units of rapidity. However,
the initial colliding beams approached with $\pm 5.5$ units of rapidity
and more than 50\% of the charged particles tracks have rapidities
outside the STAR acceptance. These particles can serve as a source of
momentum, which may quench the momentum conservation condition thus
reducing the magnitude of $\gamma_{ss}$. However, the transverse
momentum of a given track is more likely to be balanced by neighboring
particles, which have similar rapidities. This is particularly true
when considering the components of the momenta responsible for
elliptic flow. We conclude that this effect should be more essential
for lower collision energy.

The direct comparison of the momentum conservation effect (\ref{ss})
on the CME observable is presented in Fig.~\ref{TMC-cos200} for the
top RHIC energy. The total (rather than transverse) momentum
conservation is inherent in the PHSD model. In the actual calculations
the experimental acceptance $p_t>$ 200 MeV/c is taken into account; as
seen from the upper part of Fig.~\ref{TMC-cos200} the centrality
dependence of the elliptic flow $\langle v_2\rangle$ for charged
particles is rather well reproduced by PHSD. However, the experimental
same-sign correlator $\gamma_{ss}$ is underestimated substantially. We
note that the experimental acceptance essentially influences the
momentum-dependent ratio $\langle p_t\rangle^2_{acc}/\langle
p_t^2\rangle_{full}$. In reality the difference in $\gamma_{ss}$
should be even larger as discussed above. This point is in agreement
with the full HSD calculation of the hadronic background within the
CME studies in Ref.~\cite{BES-HSD}.

A similar analysis for the lower energy $\sqrt{s_{NN}}=$ 7.7 GeV is
presented in Fig.~\ref{TMC-cos200}(c). Unfortunately, measured data
for the centrality dependence of the elliptic flow are not available
at this energy but the PHSD calculated average $\langle v_2 \rangle$
for minimum bias collisions is only slightly below the
experiment~\cite{KBCTV12} due to neglecting a baryon mean-field
potential (see also the end of Sec.~\ref{ch-fl}). The calculated
correlation $\gamma_{ss}$ strongly differs from the measured values
having even the opposite sign. One should note that in this case the
same and opposite sign components are almost equal to each other ({\it
  i.e.}, there is no charge separation effect). This observation is
also nicely reproduced within the HSD model at this energy
(cf.\ Ref.~\cite{BES-HSD}).

It is of further interest to consider the average cosine of the
transverse angle difference which is independent of the reaction plane
\be \label{delta}     
\delta_{ij} \equiv
 \langle \cos(\psi_i-\psi_j)\rangle=-\frac{ \langle p_t\rangle^2_{acc}}
  {N \langle p_t^2 \ \rangle_{full}}~,
\ee
where the last equality is obtained from the transverse momentum
conservation~\cite{BKL10}. As follows from the comparison between
Eqs.~(\ref{TMC}) and (\ref{delta}), the correlator $\delta_{ij}$
differs from $\gamma_{ij}$ only by the elliptic flow coefficient $v_2$
and is expected to be more sensitive to the TMC. As one can see from
Fig.~\ref{TMC-cos200} this estimate of $\delta_{ss}$ is too large and
hardly consistent with appropriate experimental data from
Fig.~\ref{CMEm}.

Thus, the considered angular correlation $\gamma_{ss}$ is generated by
a combination of momentum conservation, which causes particles to be
preferably generated in the opposite direction, and elliptic flow
which gives more particles in the $\pm x$ direction than in the $\pm
y$ direction.  However, this source is by far not able to explain the
observed pion asymmetry in the angular correlation.  In addition, the
considered TMC is blind to the particle charge and cannot disentangle
same-sign and opposite-sign pair correlations.

\subsection{Electric charge fluctuations in the transient stage}
\label{ch-fl}

The almond-like fireball created in the early collision phase then
expands in an anisotropic way, however, the spatial anisotropy is
reduced with increasing time. In this transient stage the
electromagnetic field is strongly reduced since the spectator matter
is flying away from the formed fireball. The pressure gradients act
predominantly in the reaction plane resulting in elliptic flow $v_2$.

Strong interactions in this phase might produce significant
fluctuations in energy density (temperature), transverse momentum,
multiplicity and conserved quantities such as the net charge.  In the
plasma phase in a magnetic field an electric quadrupole can be formed
due to chiral anomaly and as a signal of that the elliptic flow
difference between $\pi^+$ and $\pi^-$ mesons is predicted
$v_2(\pi^+)< v_2(\pi^-)$~\cite{BKLY11}. Certainly, to check that the
influence of hadronic transport on observables should be taken into
account.

Furthermore, the CME~\cite{KMcLW07} predicts that in the presence of a
strong electromagnetic field at the early stage of the collision the
sphaleron transitions in a hot and dense QCD matter induce a
separation of charges along the direction of the magnetic field which
is perpendicular to the reaction $(x-z)$ plane. This charge separation
results in the formation of an electric dipole in momentum space which
breaks parity. Being interested essentially in the quark phase, we
investigate in this subsection to what extent such an electric dipole
can be generated by background statistical and electromagnetic field
fluctuations.

\begin{figure}[thb]
\includegraphics[width=0.45\textwidth,clip]{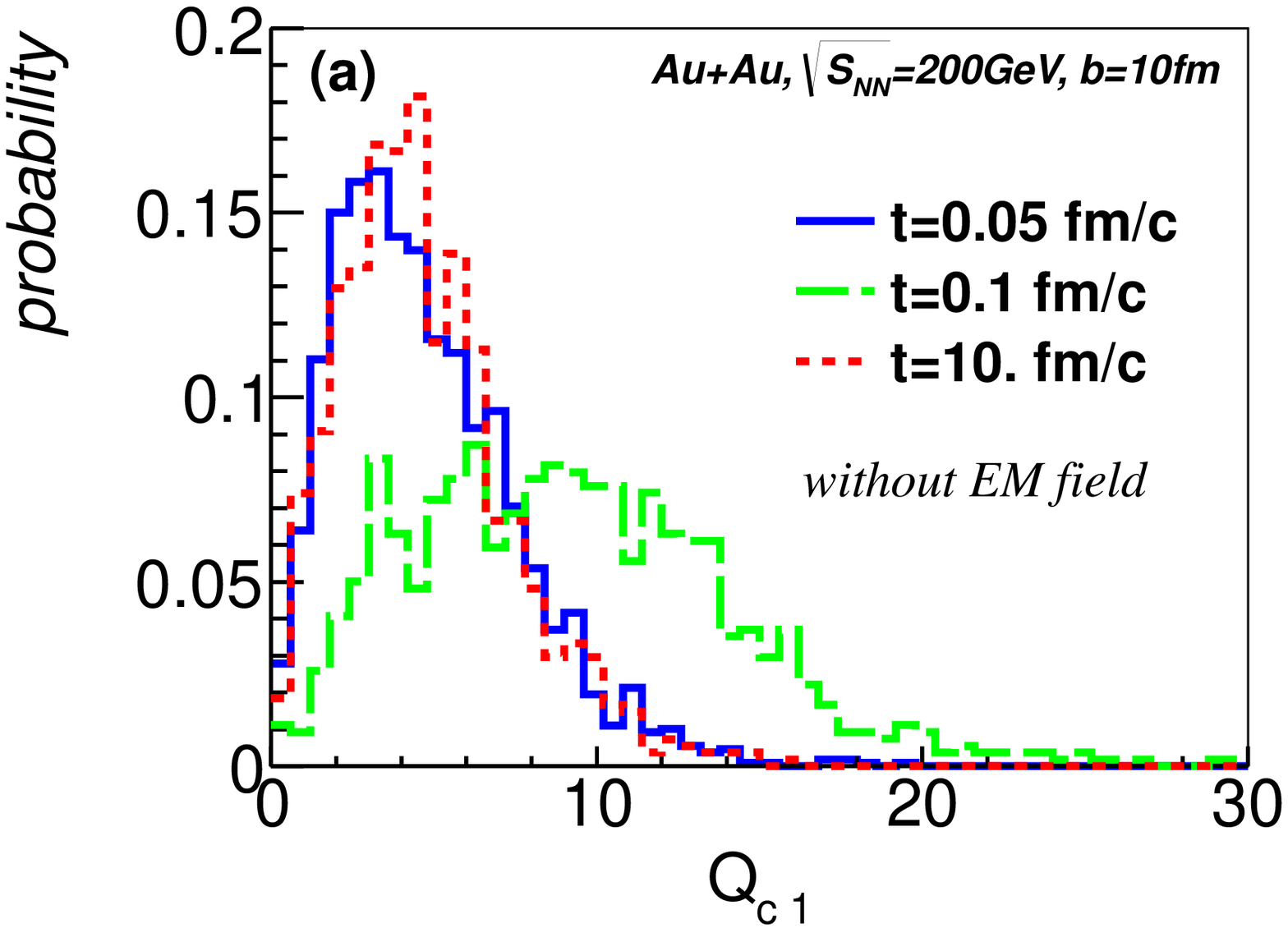}
\includegraphics[width=0.45\textwidth,clip]{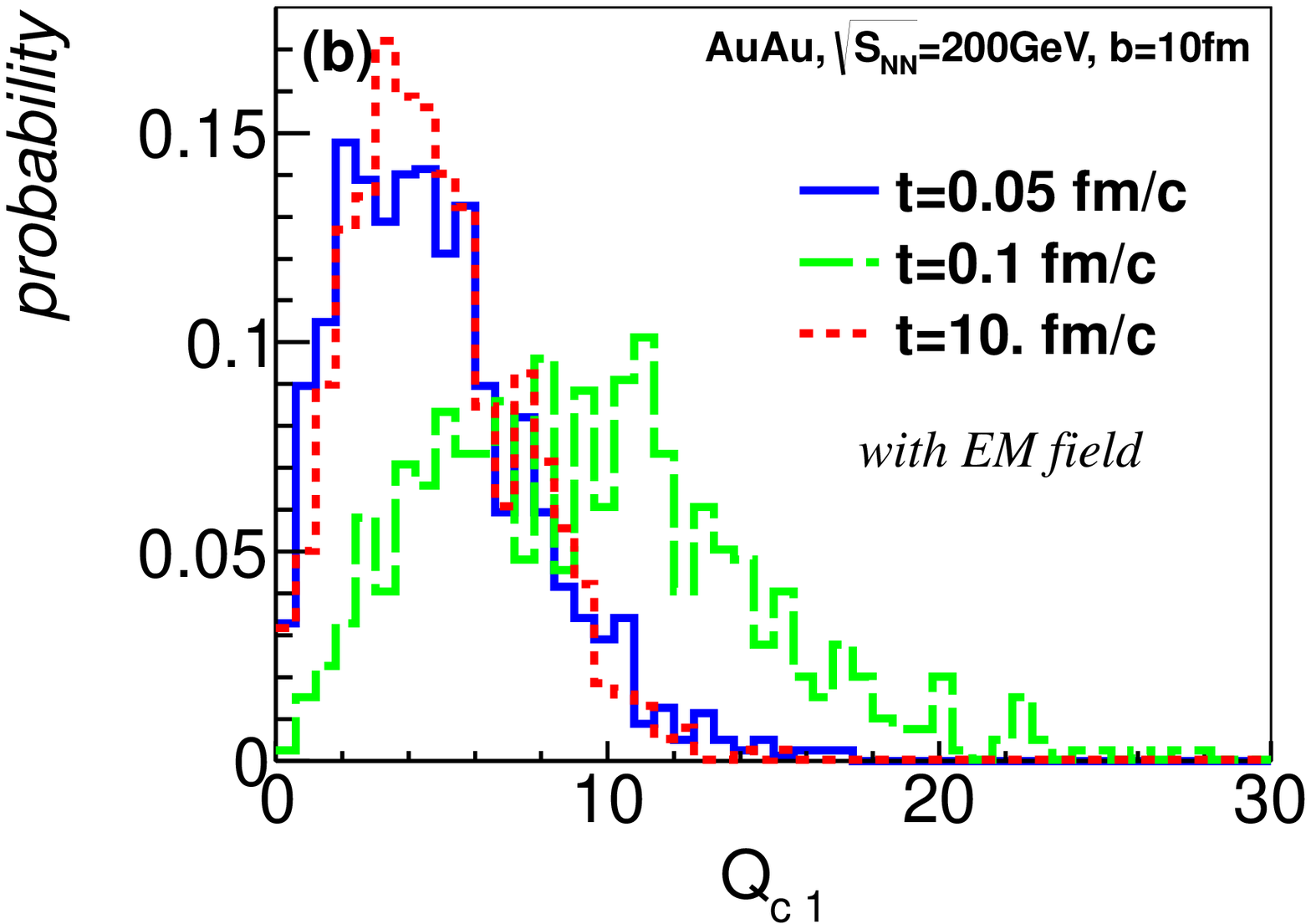}
\caption{(Color online) Probability distribution in the magnitude
  $Q_{c1}$ of the generated electric dipole at freeze-out for all
  partons. The panels (a) and (b) correspond to calculations without
  and with the electromagnetic field, respectively. The system is Pb +
  Pb at $\sqrt{s_{NN}}=$ 200 GeV at impact parameter $b=$ 10 fm.}
\label{Q-c1}
\end{figure}

Let us quantify the dipole defining the plane $\hat {Q}_{c1}$ of the
quark distribution in the transverse momentum space. The magnitude
$Q_{c1}$ and azimuthal angle $\Psi_{c1}$ of this vector can be
determined in a given event as follows:
\be
Q_{c1} \cos \Psi_{c1}=\sum_i q_i \cos \psi_i~, \nonumber \\
Q_{c1} \sin \Psi_{c1}=\sum_i q_i \sin \psi_i~,
\label{Q1}
\ee
where the summation runs over all charged particles in the event with
the electric charge $q_i$ and azimuthal angle $\psi_i$ of each
particle. Note that Eq.~(\ref{Q1}) describes the dipole shape of
charged particles (quarks or hadrons) without any reference to the
charge separation.

As seen from Fig.~\ref{Q-c1}, the average magnitude of the electric
dipole $\bar{Q}_{c1}$ at the moment of the maximum nuclear overlap
($t=$ 0.05 fm/c) is about 4 charge units with dispersion
$\sigma_{Q1}\approx 2$. At this moment the system is in the quark
phase having on average an almond-like shape and therefore is
expanding preferentially along the $x$ axis. Note that according to
Eq. (\ref{Q1}) quark net electric charges with $|q_i|<1$ are
considered. Thus, the number of quarks involved in the dipole is
large. In the next step ($t=0.1$ fm/c) of the expansion stage
(dot-dashed line in Fig.~\ref{Q-c1}) the $Q_{c1}$ distribution is
getting broader with a noticeable increase of $\bar{Q}_{c1}$. At $t=$
10 fm/c the quark-gluon phase transforms predominantly into the
hadronic phase through the dynamical coalescence mechanism and the
$Q_{c1}$ distribution becomes narrower again. The influence of the
electromagnetic field on this evolution is very weak [compare the top
  (a) and bottom (b) panels in Fig.~\ref{Q-c1}].

\begin{figure}[bt]
\includegraphics[width=0.45\textwidth,bb=50 65 390 275,clip]{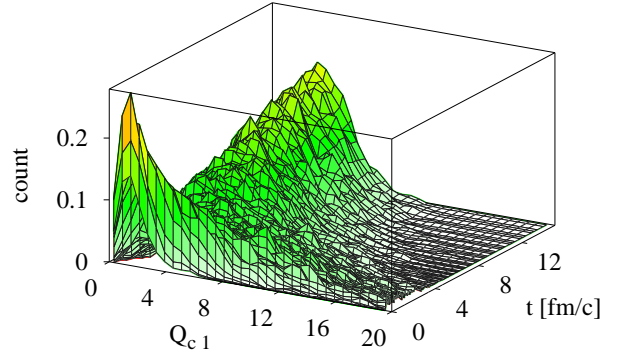}
\caption{(Color online) Electric dipole evolution of charged partons
  in $Q_{c1}-t$ presentation for Au+Au collisions at $\sqrt{s_{NN}}=$
  200 GeV with taking into account the electromagnetic field.  }
 \label{Q-c1-t}
\end{figure}

The whole evolution of the electric dipole is seen more clearly in the
3D representation in Fig.~\ref{Q-c1-t}. Indeed, the $Q_{c1}$
distribution has a pronounced peak formed shortly after the collision,
then the magnitude of the $Q_{c1}$ charge distribution is minimal
during about 3 fm/c to testify that a large-in-charge subsystem is
formed. After that the $Q_{c1}$ distribution is getting narrower
because during the expansion the $Q_{c1}$ value slightly decreases due
to parton hadronization; the maximum of the probability distribution
increases and then stabilizes after $t\approx$ 10 fm/c.

\begin{figure}[bt]
\includegraphics[width=0.45\textwidth,clip]{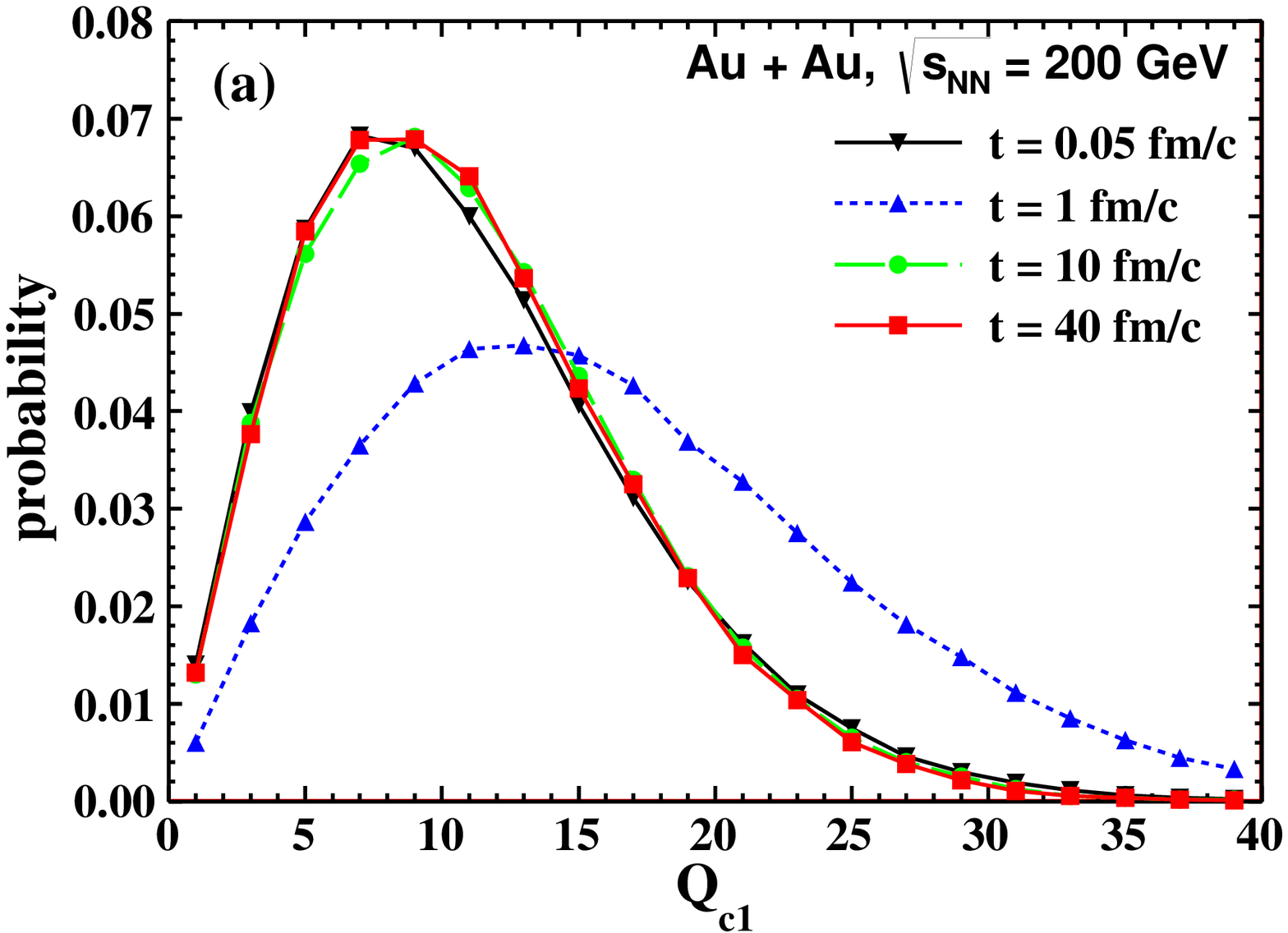}
\includegraphics[width=0.45\textwidth,clip]{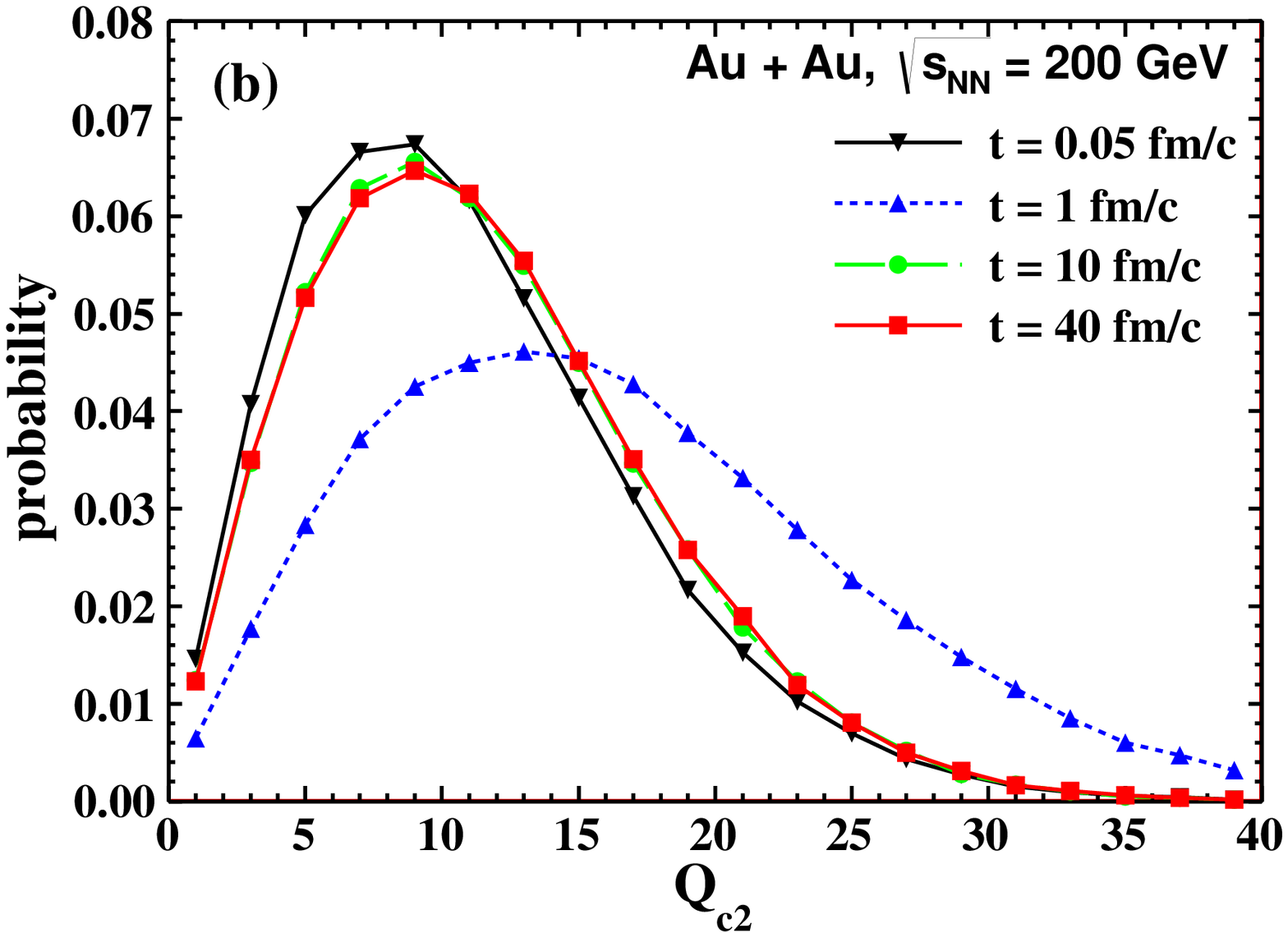}
\caption{(Color online) Distribution in the magnitude of the charge
  dipole (a) and quadrupole (b) calculated for charged particles at
  times $t=$ 0.05, 1, 10, and 40 fm/c. The system is Au + Au at
  $\sqrt{s_{NN}}=$ 200 GeV and impact parameter $b=$ 10 fm.}
\label{Q1-Q2}
\end{figure}

Similarly to the dipole, one can define a charged quadrupole formed in
heavy-ion collisions as follows:
\be
Q_{c2} \cos 2\Psi_{c2}=\sum_i q_i \cos 2\psi_i~, \nonumber \\
Q_{c2}\sin 2\Psi_{c2}=\sum_i q_i \sin 2\psi_i~.
\label{Q2}
\ee

\begin{figure}[bht]
\includegraphics[width=0.45\textwidth,clip]{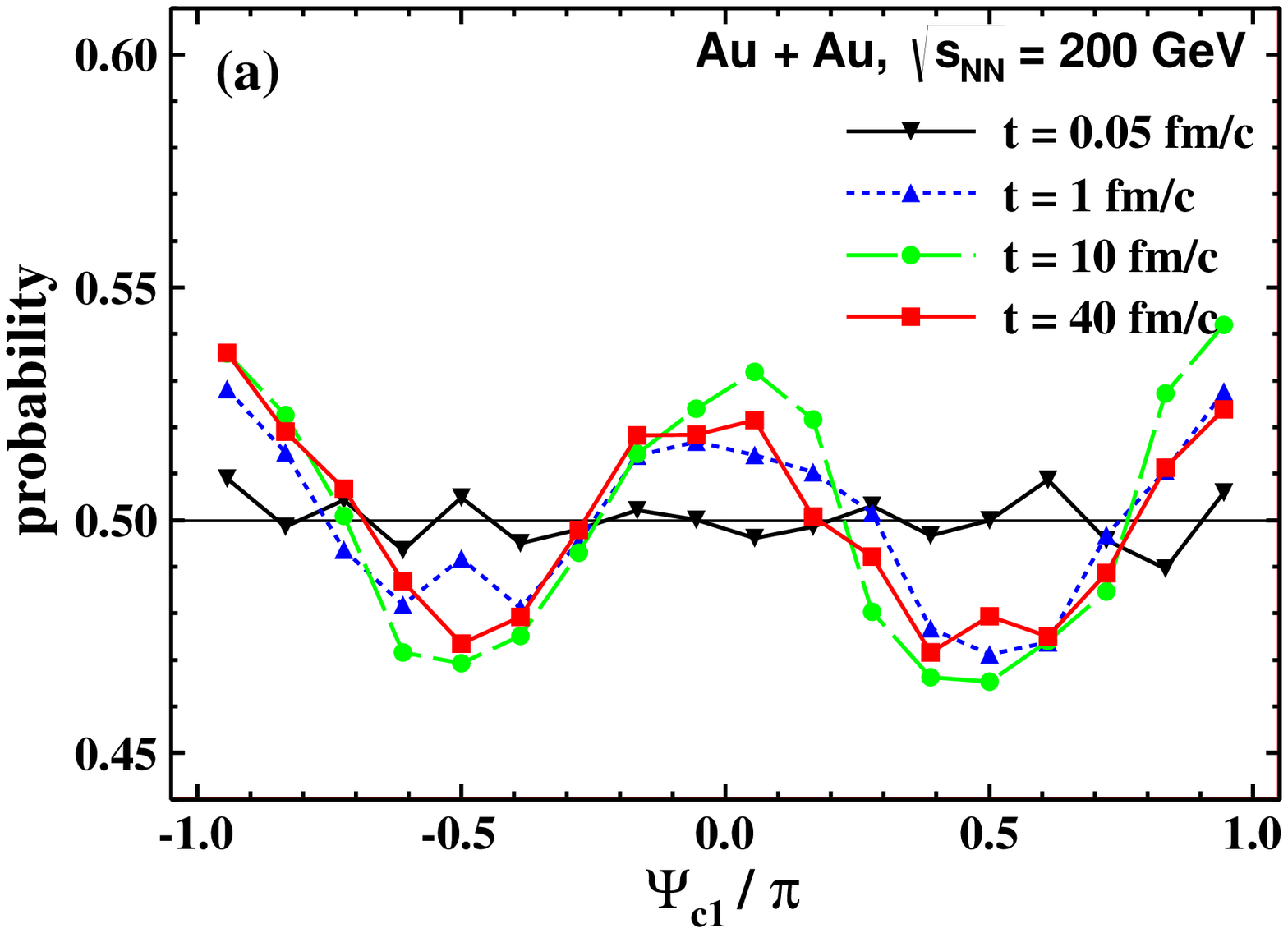}
\includegraphics[width=0.45\textwidth,clip]{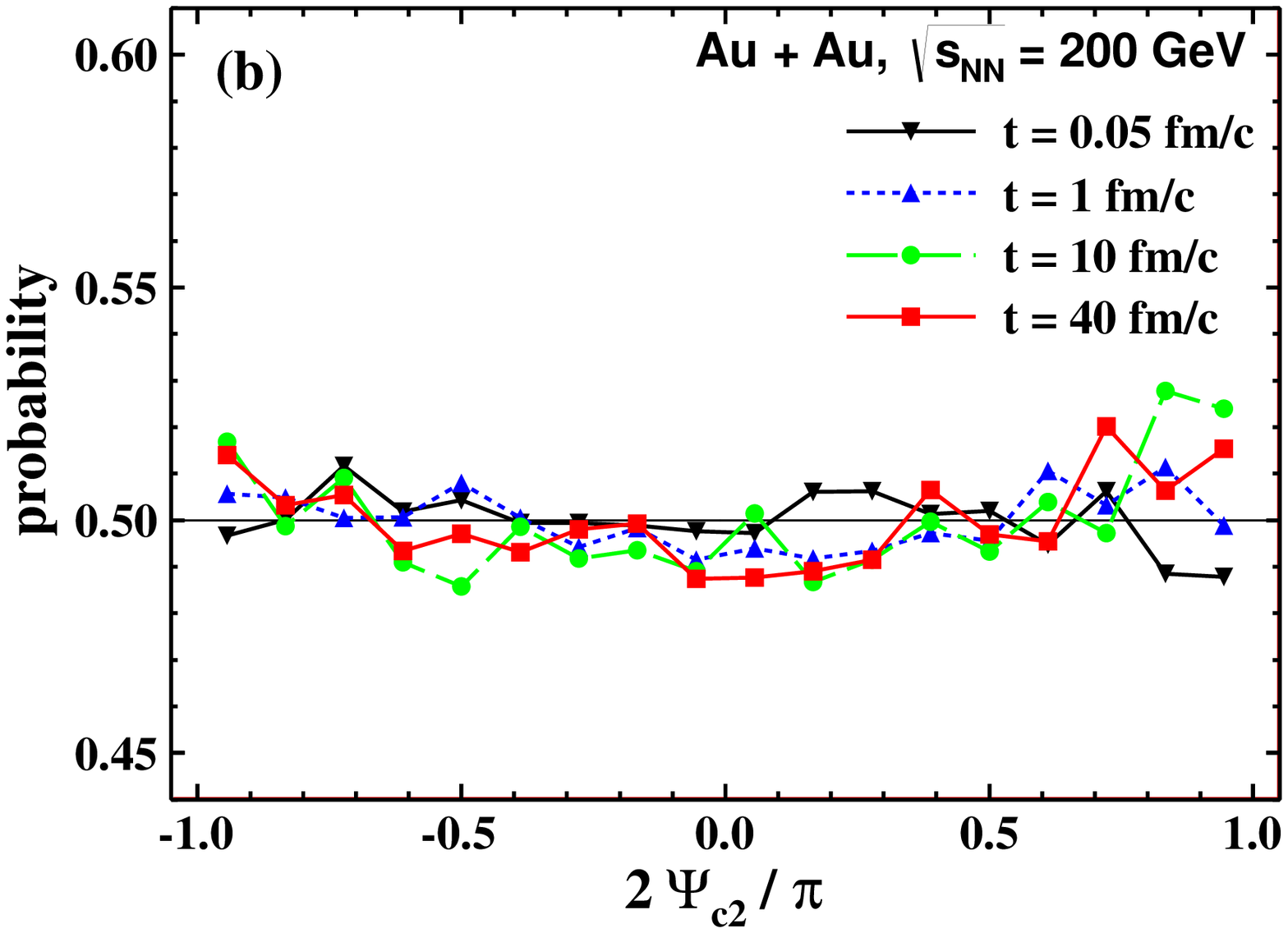}
\caption{(Color online) The same as in Fig.~\ref{Q1-Q2} but for the
  distribution in the event plane angle of the charge dipole (a) and
  quadrupole (b) at times $t=$ 0.05, 1, 10, and 40 fm/c.}
\label{Psi1-2}
\end{figure}

\begin{figure}[bht]
\includegraphics[width=0.48\textwidth,clip]{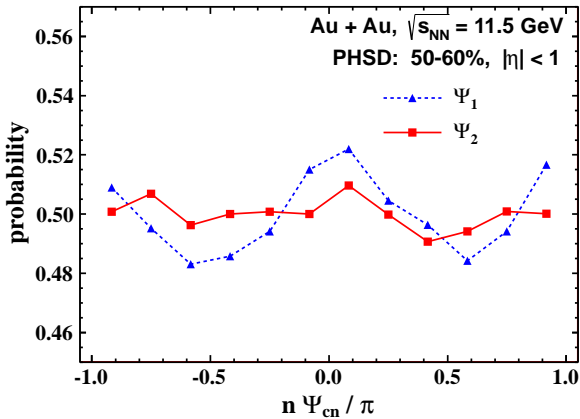}
\caption{(Color online) Angular distribution of charged particles for
  the charge quadrupole (solid line) and dipole (dashed) subsystems
  calculated for Au+Au ($\sqrt{s_{NN}}=$ 11.5 GeV) collisions.  }
\label{Qc-11}
\end{figure}

Characteristics of the time evolution of the electric dipole and
quadrupole (formed in semicentral Au+Au at $\sqrt{s_{NN}}=$ 200 GeV
collisions) for all (parton and hadrons) charged quasiparticles in the
midrapidity range are presented in the next two figures. The $Q_{c1}$
minimum observed in the pure partonic case (see Figs.~\ref{Q-c1} and
\ref{Q-c1-t}) survives also in this case. The magnitude of the $n=2$
quadrupole harmonic (presented in Fig.~\ref{Q1-Q2}) is close to that
for the dipole $n=1$, {\it i.e.}, $Q_{c2}\approx Q_{c1}$ and their
maximal values extend to values of 30--40. As is seen from
Fig.~\ref{Psi1-2}, the distributions in the reaction plane angle for
the electric quadrupole $\Psi_{c2}$ are rather flat during the whole
evolution while the electric dipole angle $\Psi_{c1}$ distribution is
flat only in the partonic phase (see $t=$ 0.05 fm/c in
Fig.~\ref{Psi1-2}) but in the hadronic phase the distribution
resembles that for the directed flow ({\it c.f.}  Fig.~\ref{Psi}). The
main axis of $\Psi_{c1}$ and $\Psi_{c2}$ can randomly be parallel or
antiparallel to the minor axis of the almond.  Like in the dipole case
(cf. Fig.~\ref{Q-c1}) the electromagnetic field has no sizable
influence on the characteristics of the electric quadrupole. When the
collision energy $\sqrt{s_{NN}}$ decreases the behavior of the dipole
and quadrupole distributions in the magnitudes $Q_{c1}$ and $Q_{c2}$
and the angle practically do not change besides some structure in the
$\Psi_{c1}$ distribution.  As seen from Fig.~\ref{Qc-11} at
$\sqrt{s_{NN}}=$ 11.5 GeV back-to-back correlations --- as specific
for the directed flow --- are manifested. This is mainly due to the
proton contribution which becomes noticeable at low collision energy.

Thus, the statistical fluctuations of ``normal'' matter in the
presence of the retarded electromagnetic field do not result in a
sizable formation of a deformed subsystem of dipole- or
quadrupole-shape during the evolution of the heavy-ion collision.

We point out, however, that such subsystems might be formed in
nontrivial topological systems due to the chiral anomaly effect. In
particular, it happens when a quark experiences both a strong magnetic
field and a topologically nontrivial gluonic field such as an
instanton~\cite{BDK12}. The inherent asymmetry --- when both instanton
and magnetic field are present --- can lead to the development of an
electric dipole moment. Physically, it can be understood as the result
of two competing effects: the spin projection produced by a magnetic
field and the chirality projection produced by an instanton
field. Such a consideration is beyond the scope of our present
microscopic study.

It is of interest that the axial anomaly in a strong external magnetic
field induces not only the CME but also the separation of the chiral
charge. The coupling of the density waves of electric and chiral
charge results in the `chiral magnetic wave' and can induce a static
quadrupole moment of the electric charge density~\cite{BKLY11}. This
chiral magnetic wave results in the degeneracy between the elliptic
flows of positive and negative pions leading to $v_2(\pi^-)>
v_2(\pi^+)$, which was estimated theoretically on the level of $\sim
30\%$ for midcentral Au+Au collisions at $\sqrt{s_{NN}}=$ 11
GeV~\cite{BKLY11}. Our PHSD calculations give about 6\% which is quite
comparable with the recently measured value of 10\%~\cite{Moh11} and
essentially smaller than the prediction of
Ref.~\cite{BKLY11}. Noteworthy that the $v_2$ degeneracy in the PHSD
version used is only due to different elastic and inelastic cross
sections for $\pi^+$ and $\pi^-$ mesons but without taking into
consideration the (small) mean-field pion-nucleus potential.  The
elliptic flow analysis of the difference between particles and
antiparticles (including kaons and baryons alongside with pions) shows
that this difference is coming mainly from the hadronic mean-field
potential~\cite{XCKL12}. Recently these $v_2$ data have been also
successfully explained in terms of a hybrid model, which combines the
fluid dynamics of a fireball evolution with a transport treatment of
the initial and final hadronic states~\cite{SKB12}. Therefore, there
is not much room for the contribution from a transient charged
quadrupole due to the chiral magnetic wave.

\subsection{Charge balance functions}

In the formation of the charged dipole and quadrupole there is no
information about a possible charge separation which could result in
an electric driving force. In principle such information can be
provided by the balance function which is based on the idea that
charge is locally conserved when particles are produced pair-wise. In
the subsequent expansion of the system and rescattering of the charge
carriers, which in principle can be hadronic or partonic, the
balancing partners are then spread out within some finite distance to
each other.  The original correlation in space-time transforms into a
correlation in momentum space in the final hadronic emission
profile. Therefore, the motion of the balancing partners suffers from
the collective expansion of the system and diffusion due to the
collisions with other particles.  The study of charge-balance
correlations hence gives insight into the production and diffusion of
charge. In particular, it is expected that the balance function is
sensitive to the delayed transition of the quark-gluon phase to a
hadronic phase~\cite{BDP00}.

\begin{figure*}[bht]
\includegraphics[width=0.45\textwidth,clip]{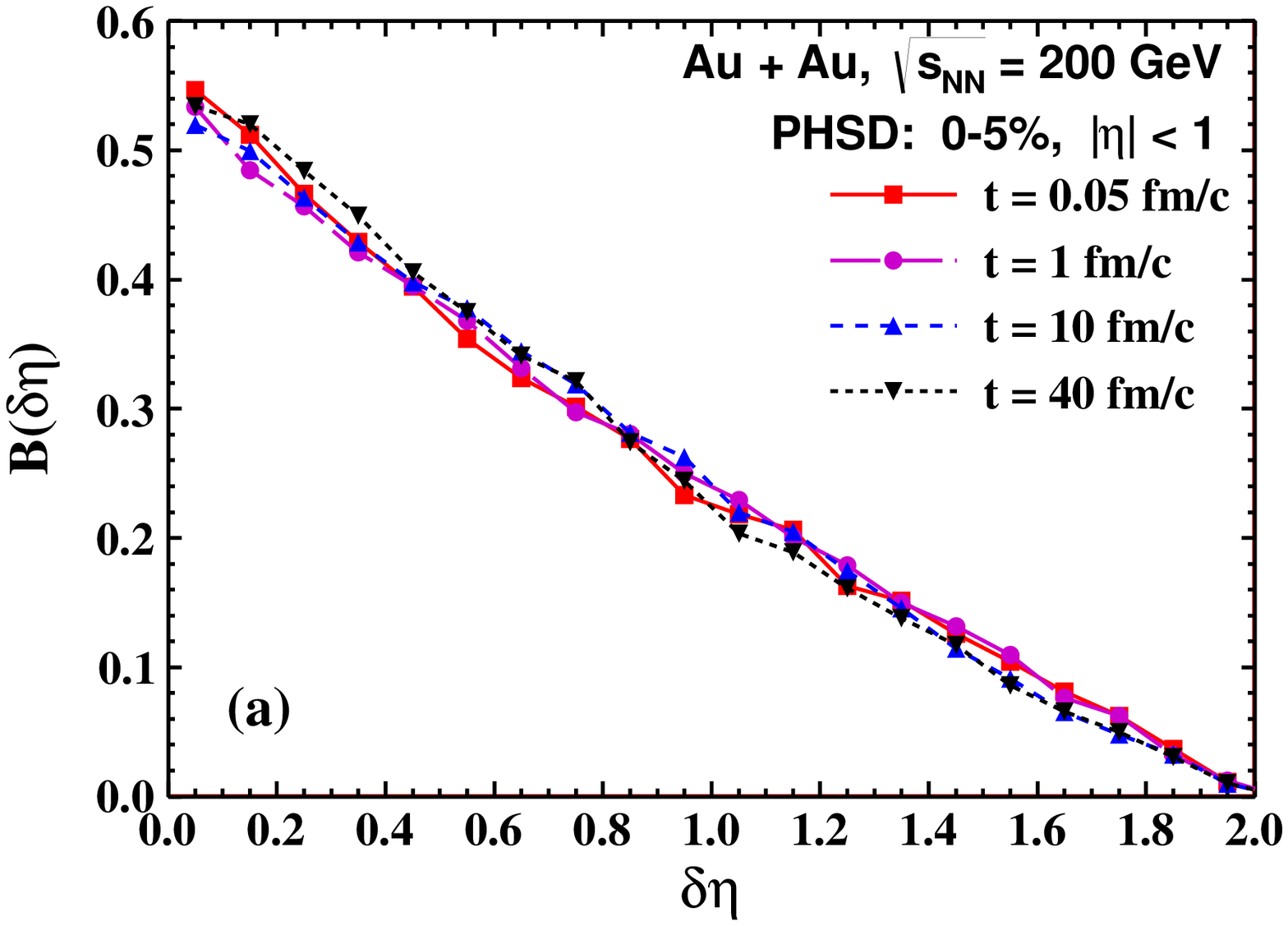}
\includegraphics[width=0.45\textwidth,clip]{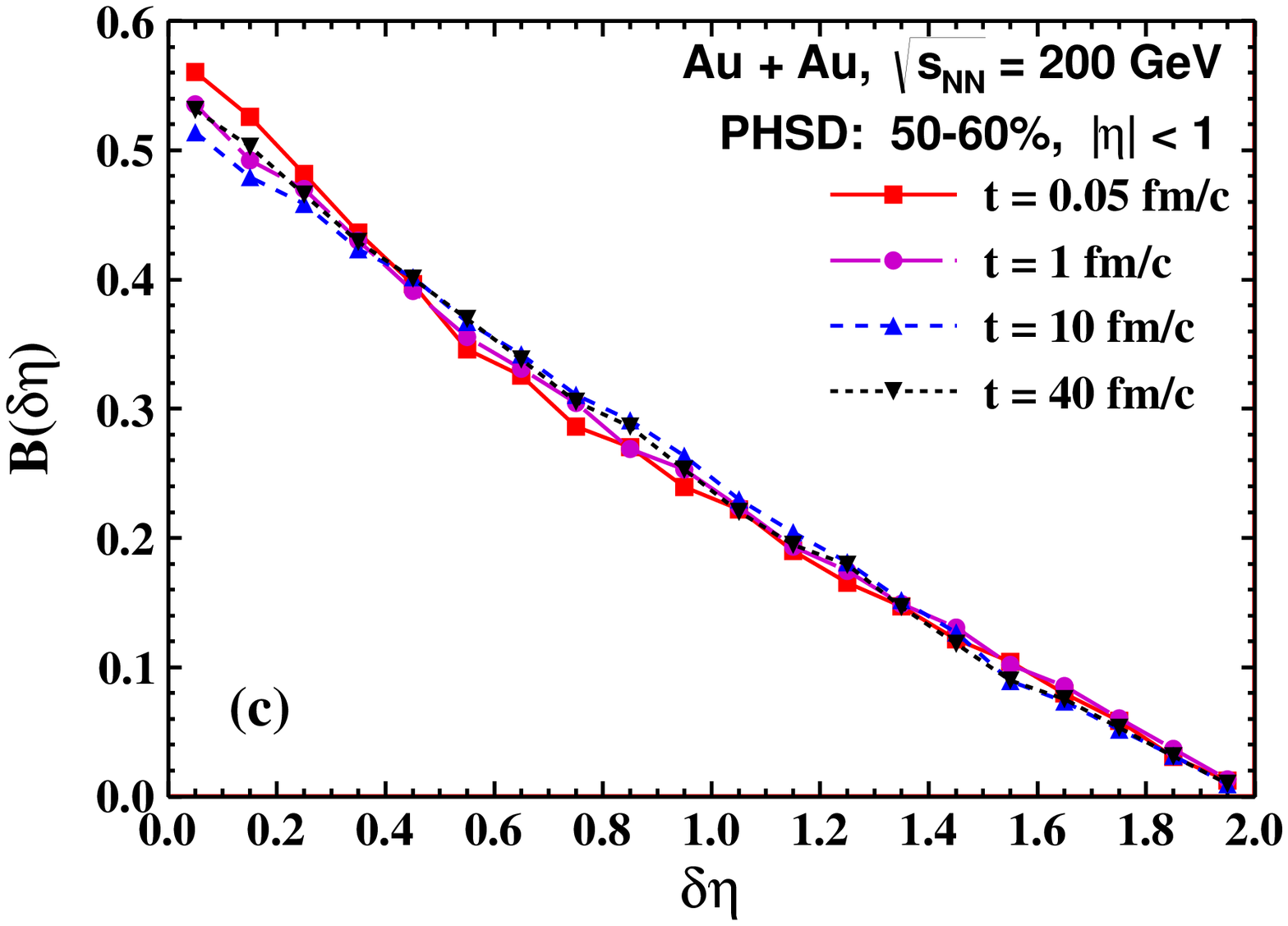}
\includegraphics[width=0.45\textwidth,clip]{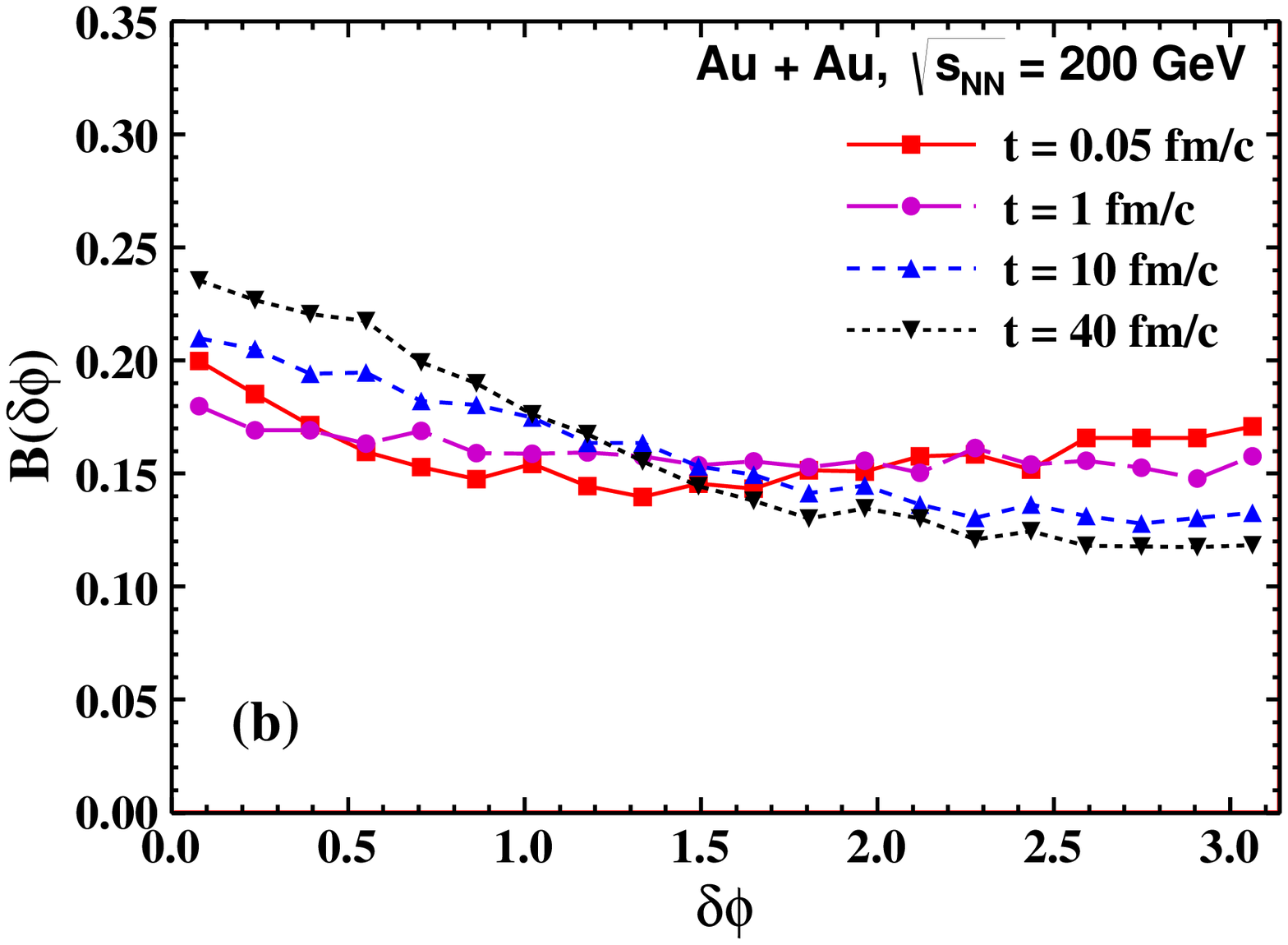}
\includegraphics[width=0.45\textwidth,clip]{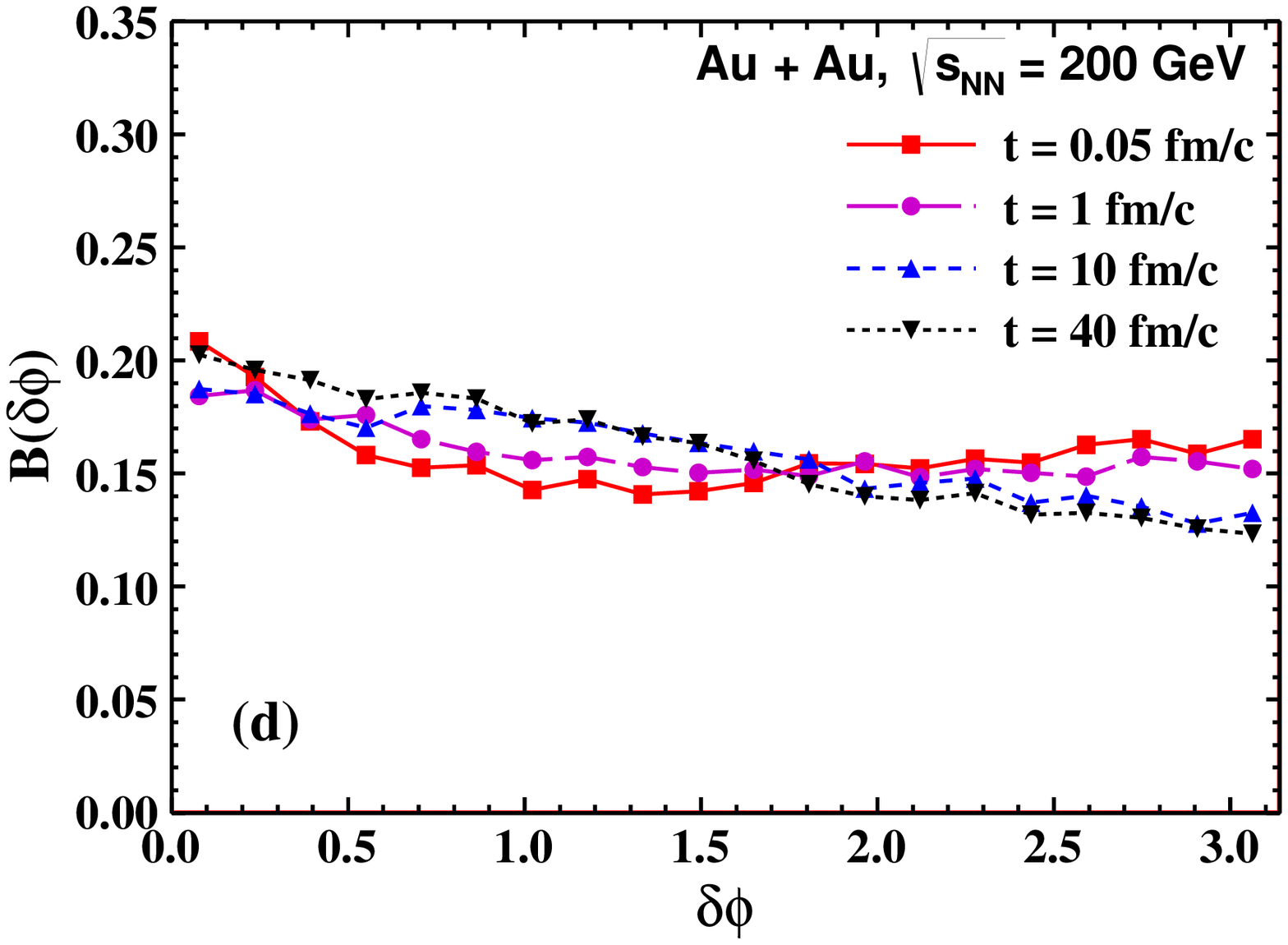}
\caption{(Color online) The balance function for charged
  quasiparticles (quark-antiquarks and $\pm$ pions) with
  $|\eta_{+/-}|<1$ from central (a),(b) and peripheral (c),(d) Au+Au
  ($\sqrt{s_{NN}}=$ 200 GeV) collisions at $t=$ 0.05, 1, 10, and 40
  fm/c.}
 \label{BF-t}
\end{figure*}

Thus, whenever a positive charge is created, a negative charge arises
from the same point in space-time and both particles then tend to be
focused in the same rapidity and azimuthal angle by collective
flow. This results in a correlation between positive and negative
charges, {\it i.e.}, for every positively charged particle emitted at
an angle $\psi_+$, there tends to be a negatively charged particle
emitted with $\psi^-\approx\psi^+$ and similar rapidity. Charge
balance functions \cite{BDP00} represent a measure of such
correlations, and have already been investigated as a function of
relative rapidity for identified particles and for relative
pseudorapidity $\eta$ for nonidentified
particles~\cite{Aggarwal:2010ya,STAR-BF03}. Generally, the balance
function $B(p_a|p_b)$ is a six-dimensional function of the particle
momenta. In the context of studies of the separation of balancing
charges, the discussion is reduced to the difference $({\bf p}_1 -
{\bf p}_2)$. In particular we will focus on the charge balance
function in relative pseudorapidity $\delta\eta$, {\it i.e.},
$B(p_a|p_b) \to B(\delta \eta,\eta_w)$ and similarly for the azimuthal
angle $\psi$~\cite{Bo05}.

Charge balance functions are constructed in such a way that like-sign
subtractions statistically isolate the charge balancing partners,
 \be \label{bal.f}
B(\delta \eta,\eta_w)&=& \frac{1}{2} \left( \
\frac{N_{+-}(\delta\eta,\eta_w)-N_{++}(\delta\eta,\eta_w)}{N_+}
\right. \nonumber
\\ &+&\left.
\frac{N_{-+}(\delta\eta,\eta_w)-N_{--}(\delta\eta,\eta_w)}{N_-} \
\right),
 \ee
where the conditional probability $N_{+-}(\delta\eta,\eta_w )$ counts
pairs with opposite charge which satisfy the criteria that their
relative pseudorapidity $\delta\eta=\eta_+-\eta_-$ in a given
pseudorapidity window is $\eta_w$, $(\delta\eta \in \eta_w)$, whereas
$N_+ (N_-)$ is the number of positive (negative) particles in the same
interval.  Similarly for $N_{++}$, $N_{--}$, and $N_{-+}$. The factor 1/2
ensures the normalization of $B(\delta \eta, \eta_w)$. All terms in
Eq. (\ref{bal.f}) are calculated within PHSD using pairs from a given
event and the resulting distributions are summed over all events.

Both balance-function and charge-fluctuation observables are generated
from one-body and two-body observables which necessitates that they
may be expressed in terms of spectra and two-particle correlation
functions. The charge fluctuation is a global measure of the charge
correlation and the balance function is a differential measure of the
charge correlation; it therefore carries more information. Writing
$N_{\pm}= \langle N_{\pm}\rangle_{\eta_w}+\delta N_\pm$, where
$\langle \dots \rangle_{\eta_w}$ denotes the average in the phase-space
region $\eta_w$, it is easy to show \cite{JP02} that
\be \label{dif}
\frac{\langle(Q_{ch} - \langle Q_{ch}\rangle)^2\rangle}{\langle N_{ch}\rangle}
\simeq 1 - \int_0^{\eta_w} d\delta \eta \ B(\delta \eta|\eta_w ) \ , 
\ee
where $Q_{ch} = N_+ - N_-$ and $N_{ch} = N_+ + N_-$.

From this example, one can readily understand how balance functions
identify balancing charges. For any positive charge, there exists only
one negatively charged particle whose negative charge originates from
the point at which the positive charge was created.  By subtracting
from the numerator the same object created with positive-positive
pairs, one is effectively subtracting the uncorrelated contribution
from the distribution.

It is expected that charge balance functions are sensitive to the
separation of balancing charges in momentum space and give insight
into the dynamics of hadronization~\cite{BDP00}. Indeed, such a pair
is composed of a positive and negative particle (or particle and
antiparticle) whose charge originates from the same point in
space-time.  According to~\cite{BDP00}, if a quark-gluon plasma
results in a large production of new charges (quark-antiquark pairs)
late in the reaction, a tight correlation between the balancing
charge-anticharge pairs would provide evidence for the creation of
this novel state of matter.

The time evolution of the $\delta \eta$- and $\delta\phi$-dependent
charge balance function is demonstrated in Fig.~\ref{BF-t} for central
[(a),(b)] and peripheral [(c),(d)] Au+Au collisions at
$\sqrt{s_{NN}}=$ 200 GeV. The times $t=$ 0.05 and 1 fm/c correspond to
the developed quark phase which ends at about $t=$ 10 fm/c
(cf.\ Fig.~\ref{Dp-cromo-comp}) while at $t=$ 40 fm/c the system is in
a purely hadronic phase. We recall that hadronization in the PHSD
model is realized via a crossover transition and quasiparticle
rescattering is included in both the partonic and hadronic phase.

As follows from Fig.~\ref{BF-t}, there is a very small difference in
the time evolution of charge balance function $B(\delta \eta)$ at both
centralities. As to $B(\delta \phi)$ a rather clear enhancement is
seen for central collisions at $\delta\eta\sim\delta\phi\sim 0$ while
this dependence is essentially weaker for peripheral collisions. This
observation is in qualitative agreement with experiment but
enhancement effect is too small.  Thus the expectation of a high
sensitivity of the balance function to the hadron phase transition and
particle diffusion seems somewhat too optimistic.

\begin{figure}[bht]
\includegraphics[width=0.45\textwidth,clip]{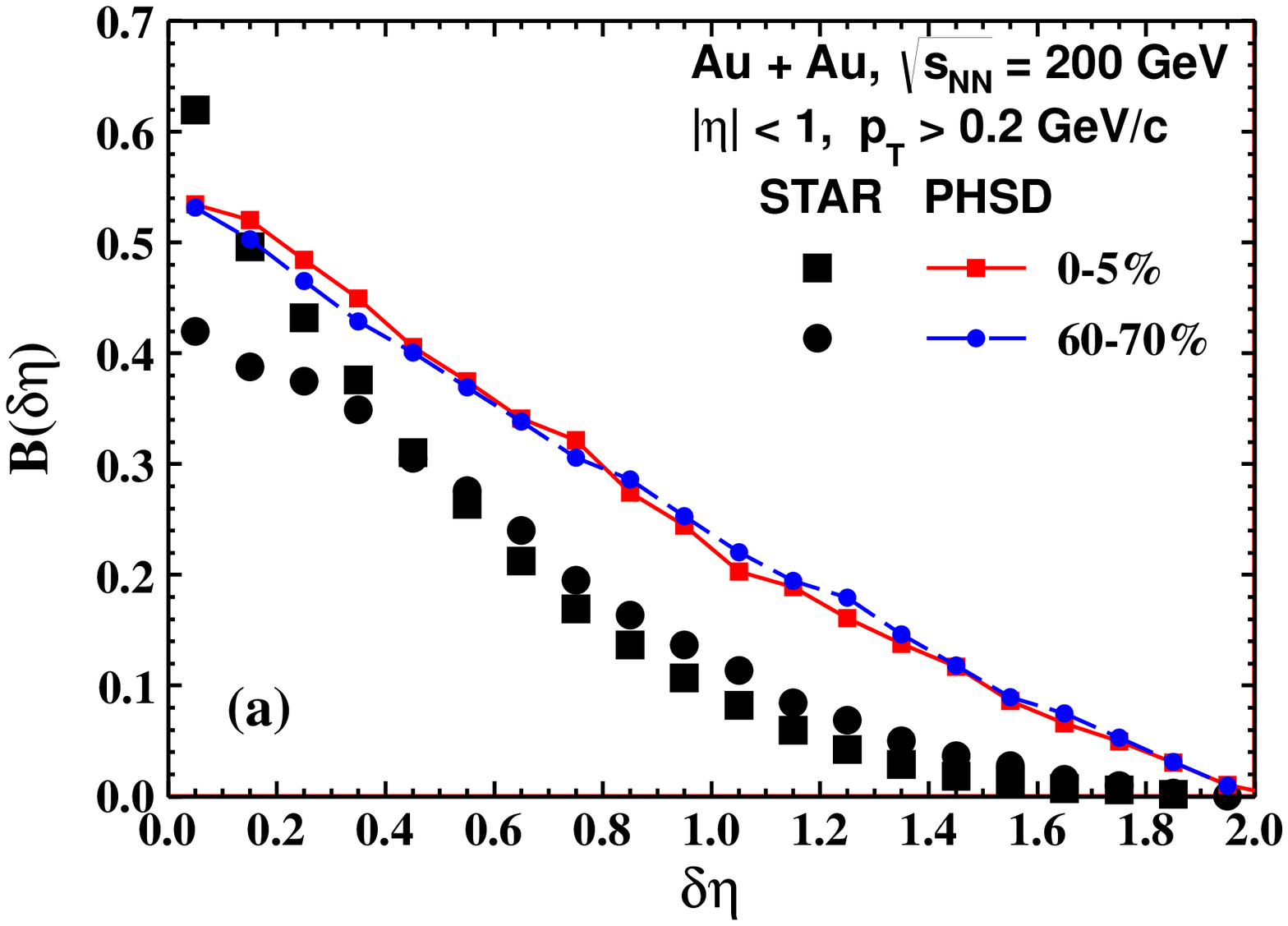}
\includegraphics[width=0.45\textwidth,clip]{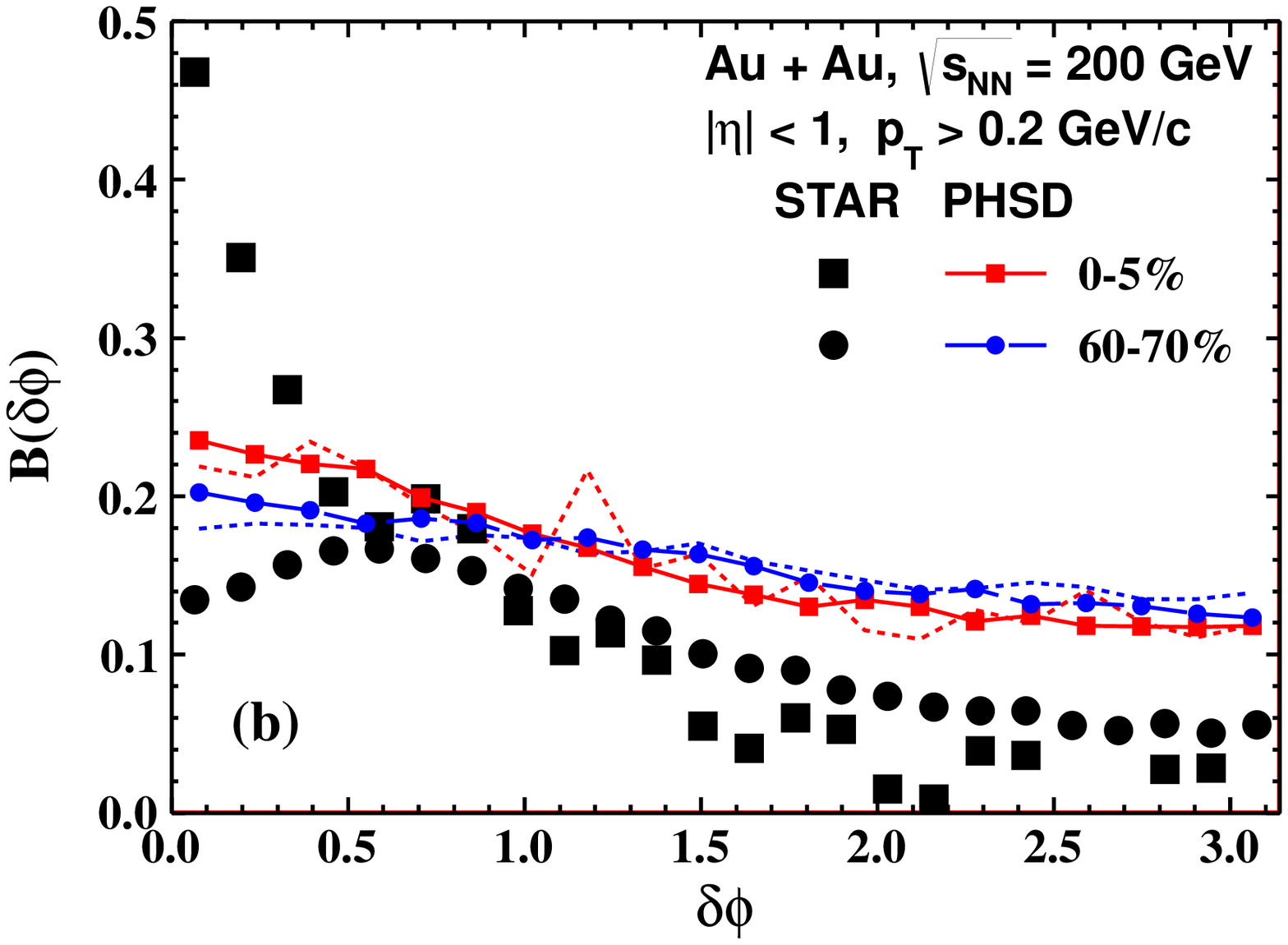}
\caption{(Color online) The balance function for pseudorapidity (a)
  and azimuthal angle (b) of charged pions with $|\eta_{+/-}|<1$ and
  $p_t>$ 0.2 GeV/c from central and peripheral Au+Au collisions at
  $\sqrt{s_{NN}}=$ 200 GeV. The dotted lines for $\delta \phi$
  correspond to calculations including the electromagnetic field
  effects. The experimental data points are from
  Ref.~\cite{Aggarwal:2010ya}.}
\label{BF}
\end{figure}

The direct comparison with experiment of the charge balance function
is presented in Fig.~\ref{BF} for Au+Au collisions at $\sqrt{s_{NN}}=$
200 GeV calculated within PHSD for central and peripheral ($b=$ 10 fm)
collisions. Here the charge conservation law is locally fulfilled in
each quasiparticle collision. Hadronic resonance decays are taken into
account in PHSD but corrections due to final state interactions for
small $\delta\phi$ are computationally very involved (and uncertain).

As is seen from Fig.~\ref{BF} there is a maximum for $\delta \eta \sim
\delta \phi \sim 0$ but no large difference is observed for different
centralities (apart from the $\delta \phi$ distribution). This finding
is in agreement with the kinetic results of UrQMD and HIJING model
calculations~\cite{Aggarwal:2010ya} of the width of the balance
function in terms of $\delta \eta$ which also show no narrowing of the
peak for central collisions as observed in experiments. The inclusion
of the electromagnetic field [the dotted line in Fig.~\ref{BF}(b) for
  $b=$ 10 fm] practically shows no influence and does not result in
any additional focusing in central collisions.

We find that PHSD does not describe quantitatively the experimental
balance functions. There are some claims that the blast-wave model can
resolve this discrepancy~\cite{Pr10-1,Pr10,Aggarwal:2010ya}. The
blast-wave model is in fact a parametrization of the kinetic
freeze-out configuration motivated by a hydrodynamical model for the
system described in local thermal equilibrium. The system is then
completely characterized by the collective velocity profile,
freeze-out temperature, and the freeze-out surface which is usually
associated with some volume. Generally, the blast-wave model
parameters may be varied in a large parameter space to fit
experimental data. These single-particle freeze-out properties can,
{\it e.g.}, be parametrized as suggested in \cite{BW-STAR05} to study,
for example, the evolution of flow. However, the change in the kinetic
freeze-out temperature and the increase of collective flow alone fail
to explain the observed focusing of the balance function for more
central collisions~\cite{Pr10}.

With regard to charge-balance correlations, the blast-wave model needs
additionally to incorporate local charge conservation. This can be
achieved in the following way: Instead of generating a single particle
at a time, an ensemble of particles with exactly conserved charges is
generated in such a way to remain unchanged the single-particle
distributions. For the relative distribution of the pairs within an
ensemble, a Gaussian distribution is assumed with dispersions
$\sigma^2_\eta$ and $\sigma^2_\phi$ for rapidity and transverse angle,
respectively. Treating these dispersions as free parameters at every
centrality it is possible to tune the narrowing effect for central
events~\cite{Pr10}. It is of interest that at the exactly central
collision $\sigma^2_\eta=\sigma^2_\phi=0$ and they strongly grow with
impact parameter reaching $\sigma^2_\eta\approx 0.6$ and
$\sigma^2_\phi/\pi\approx 0.4$ for centralities about
70\%~\cite{Pr10}. However, the additional assumption that balancing
charges at the freeze-out are strongly correlated in momentum space is
in conflict with the basic model assumption on thermodynamic
equilibrium of the system.

In Ref.~\cite{Pr10} the charge balance function was applied for the
analysis of the CME. The charge separation between opposite-charge and
same-charge two-pion correlators $\gamma_P$ was defined as
\be
\label{pr}
 \gamma_P \equiv \frac{1}{2} (2 \gamma_{+-}- \gamma_{++}- \gamma_{--})
 =  \gamma_{+-}- \gamma_{ss}~,
\ee
where the angle brackets in Eq.~(\ref{cos}) include the balance
function $B(p_+|p_-)$ as a weight factor for the balancing
charges. The quantity $\gamma_P$ can be estimated from available
experimental data~\cite{STAR-CME}. Since the PHSD is not successful in
reproducing the charge balance function, there is not much sense to
apply it for the charge separation $\gamma_P$. As follows from the
comparison between the STAR data and the blast-wave model (including
correlations and rescaling) results in reproducing the experimental
normalization; the charge balance correlations for the relativistic
charge separation are of the same size as the experimental signal and
exhibit a similar qualitative behavior with respect to the centrality
dependence~\cite{Pr10}.  The authors of Ref.~\cite{Pr10} claim that
their results are solid on the level of 10--20\%.  The calculation of
uncertainties originates predominantly from the particular
parametrization of both the blast-wave model itself and, in
particular, of the centrality dependence of the charge separation
$B(p_a|p_b)$ in the azimuthal angle. However, the considered breakup
physics differs significantly from more realistic scenarios as has
been shown recently in Ref.~\cite{Kumar:2012xa}; the freeze-out
temperature and baryon chemical potential --- defining the chemical
composition of the system --- noticeably depend on
centrality. Furthermore, there is some inconsistency in using a
reaction-plane independent fit of the balance function for the CME
signal where azimuthal angles are measured with respect to the
reaction plane. It is also unclear how the extracted parameters change
with the collision energy, however, the first preliminary STAR data on
the collision-energy dependence of the balance function have been
published recently~\cite{Wa11}.  Thus, a further careful study of this
issue is needed.

\section{The CME observable}
\label{Sec:Observable}

\begin{figure*}[thb]
\includegraphics[width=0.45\textwidth] {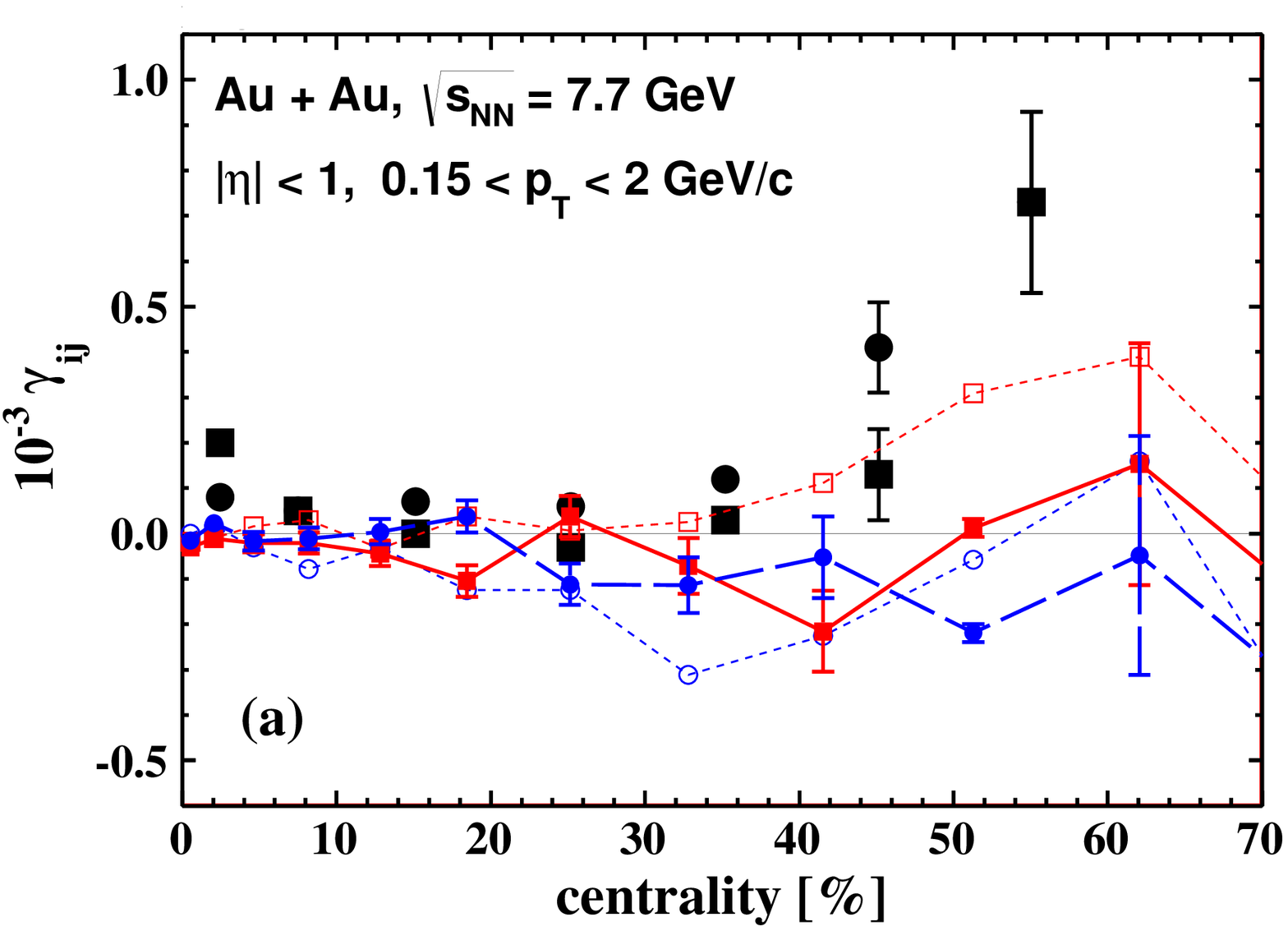}
\includegraphics[width=0.45\textwidth] {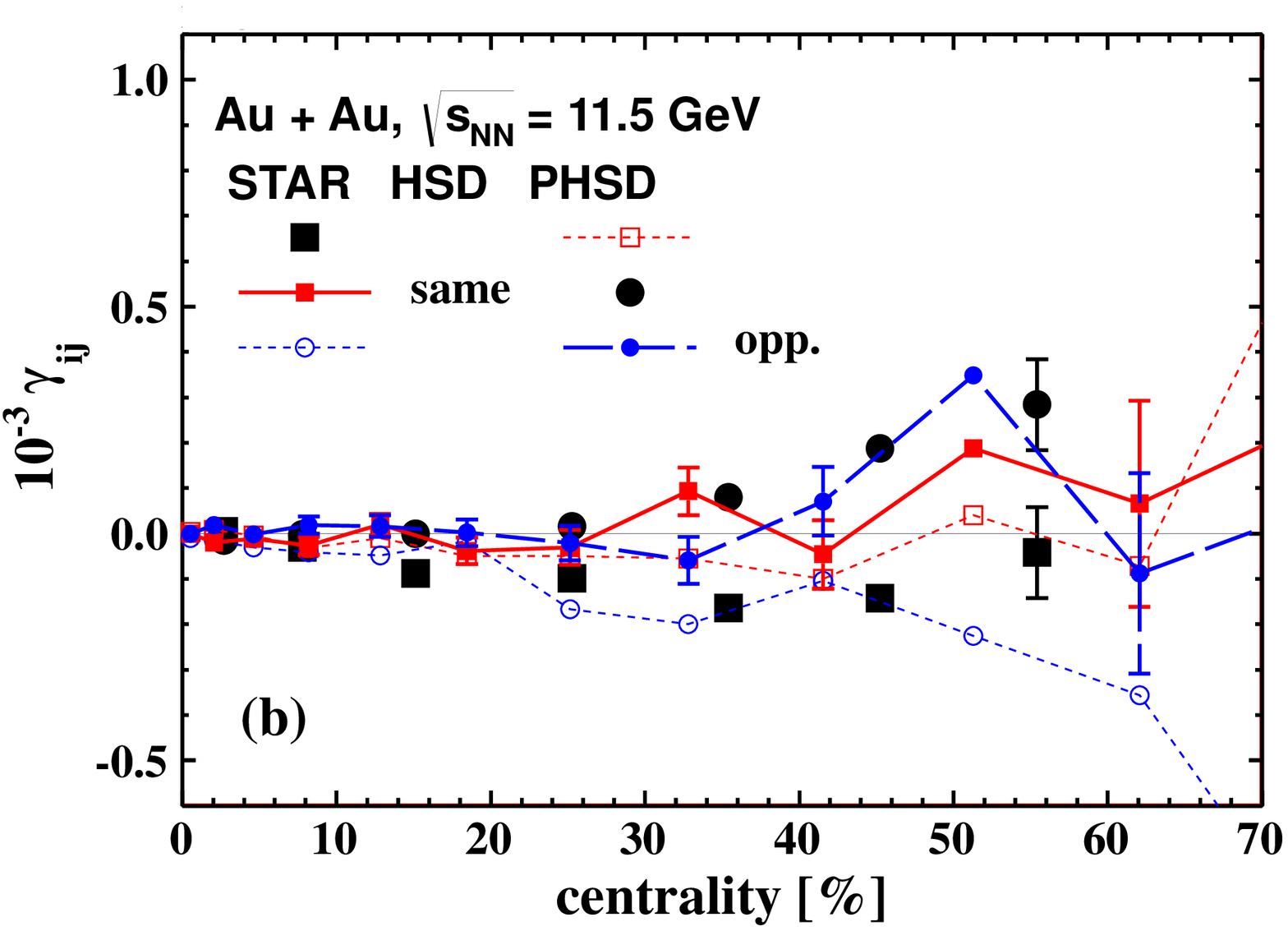}
\includegraphics[width=0.45\textwidth] {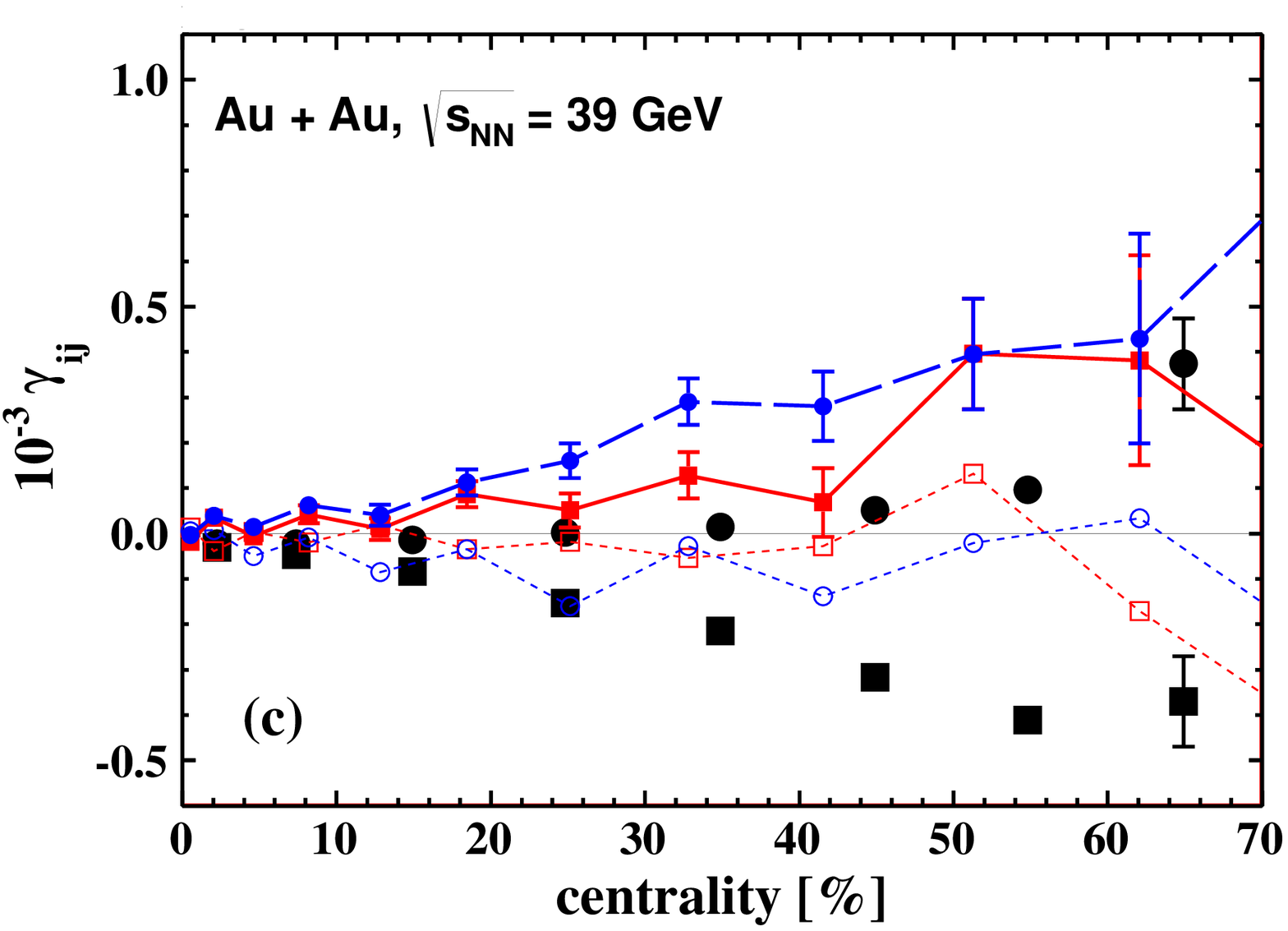}
\includegraphics[width=0.45\textwidth] {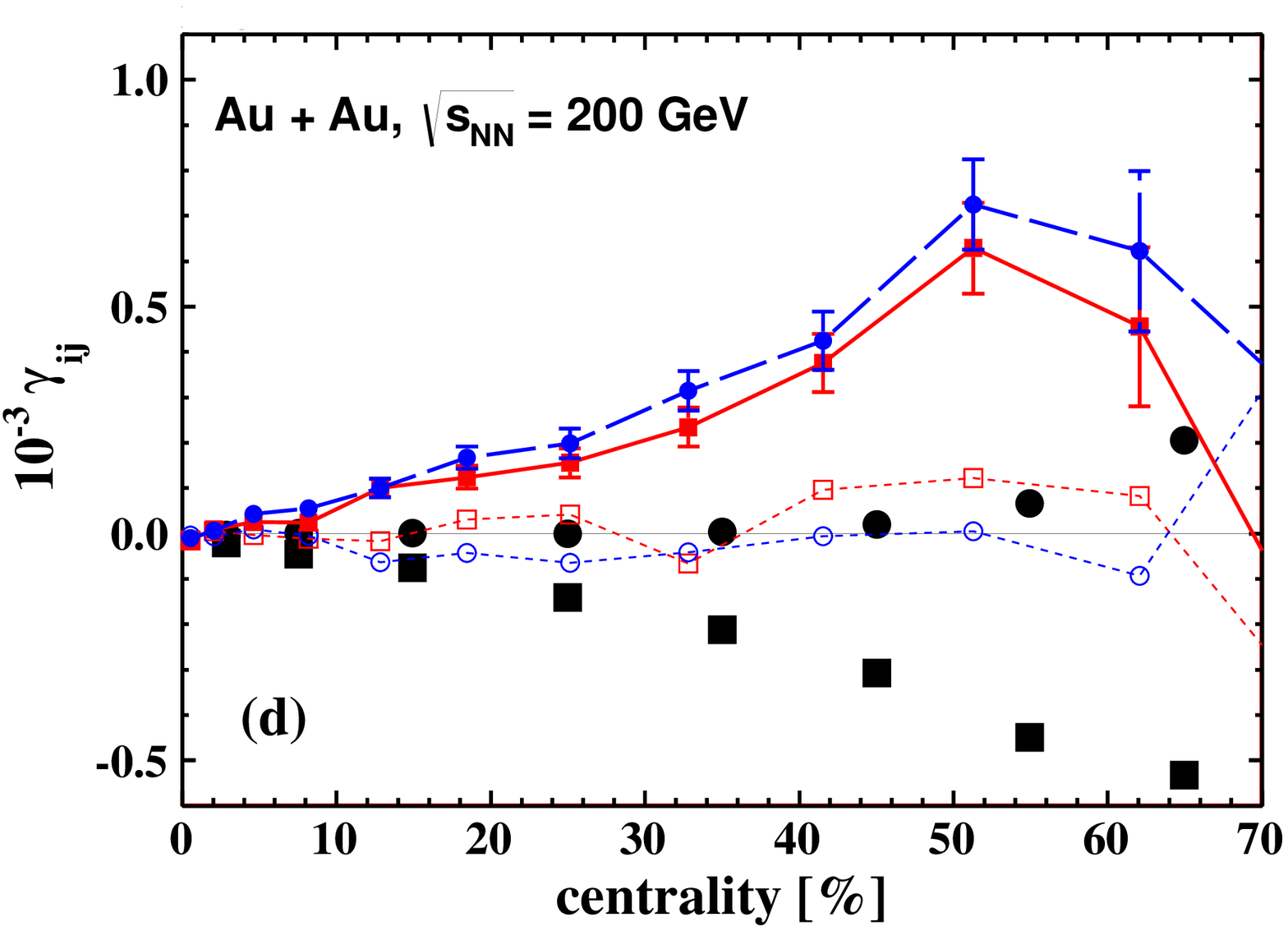}
\caption{(Color online) Angular correlations of opposite- and
  same-charge pions in azimuthal angle for Au+Au collisions at
  $\sqrt{s_{NN}}=$ 7.7, 11.5, 39, and 200 GeV as a function of
  centrality. The full symbols are preliminary STAR data~\cite{BES11}
  as well as published STAR data for $\sqrt{s_{NN}}=$ 200
  GeV~\cite{STAR-CME}.  }
\label{C2Hpr}
\end{figure*}

The experimental signal of the possible CME is the azimuthal angle
correlator calculated according to Eq.~(\ref{cos}). The experimental
acceptance $|\eta|<1$ and 0.20 $<p_t<2$ GeV has been also incorporated
in the theoretical PHSD calculations. Note that the theoretical
reaction plane is fixed exactly by the initial conditions rather than
by a correlation with a third charged particle as in the
experiment~\cite{BES11}. Thus, within PHSD we calculate the observable
(\ref{cos}) as a function of the impact parameter $b$ or the
centrality of the nuclear collisions which should be considered as a
background of the CME signal. A comparison of the measured angular
correlator with result of calculations is presented in
Fig.~\ref{C2Hpr}. We mention that the calculation of this correlation
is a very CP time consuming process and the proper statistical error
bars are shown in Fig.~\ref{C2Hpr}.

At the lowest measured energy $\sqrt{s_{NN}}=$ 7.7 GeV the results for
oppositely and same-charged pions are very close to each other and
show some enhancement in very peripheral collisions. The centrality
distributions of $\gamma_{ij} $ are reasonably reproduced by the PHSD
and HSD calculations presented in the same picture. Note that the
scalar quark potential is not zero at this low energy but absent in
the HSD model. The striking result is that the case of
$\sqrt{s_{NN}}=$ 7.7 GeV drastically differs from $\sqrt{s_{NN}}=$ 200
GeV [cf.\ panels (a) and (d) in Fig.~\ref{C2Hpr}]. The picture
quantitatively changes only slightly when one proceeds to
$\sqrt{s_{NN}}=$ 11.5 GeV [see the panel (b) in Fig.~\ref{C2Hpr}]
though the value of $\gamma_{ij}$ at the maximum (centrality 70\%)
decreases a little bit in the calculations. Experimental points at
larger centrality are not available but are of great interest. In
addition, one may indicate a weak charge separation effect in the
experimental data because statistical error bars are very small (less
than the symbol size). Unfortunately, the calculated error bars are
rather large to specify the charge separation effect. The influence of
the electromagnetic field here is negligible. The calculated and
measured correlation functions for oppositely and same charged pions
are shown in Fig.~\ref{C2Hpr} for the available three BES
energies. The case for the top RHIC energy $\sqrt{s_{NN}}=$ 200 GeV is
also presented for comparison.

If one looks now at the results for $\sqrt{s_{NN}}=$ 39 GeV, the
measured same- and oppositely charged pion lines are clearly
separated, being positive for the same-charged and negative for the
oppositely charged pions to be strongly suppressed. The PHSD model is
not able to describe this picture and overestimates the data with
increasing energy. These growing large values of $\gamma_{ij}$ are due
to the scalar parton potential which increases with the collision
energy. The HSD version predicts a very small effect in qualitative
agreement with our earlier analysis~\cite{BES-HSD}. Though both models
provide the charge separation essentially smaller than the measured
one, the PHSD has a satisfying feature: the same-charge points are
above the oppositely charged ones to be in agreement with
experiment. The same situation is observed in the case of
$\sqrt{s_{NN}}=$ 200 GeV; a small difference between them is seen in
very peripheral collisions: the oppositely charged correlation jumps
to zero at centrality $\sim$ 70\% for $\sqrt{s_{NN}}=$ 39 GeV while
corresponding data at 200 GeV are not available.

\begin{figure}[tb]
\includegraphics[width=0.42\textwidth,clip]{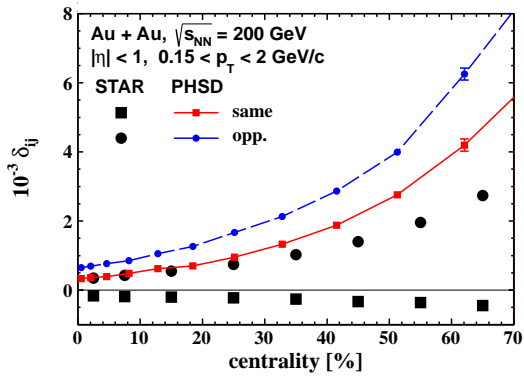}
\caption{(Color online) Angular correlations of opposite- and
  same-charge pions for the cosine of the difference in the azimuthal
  angles for Au+Au collisions at $\sqrt{s_{NN}}=$ 200 GeV as a
  function of centrality. The experimental data points are
  from~\cite{STAR-CME}.}
\label{CMEm}
\end{figure}

Though the results at $\sqrt{s_{NN}}=$ 7.7 and 11.5 GeV roughly can be
considered as a background of the CME, at higher energies it is
impossible to identify the true effect of the local parity violation
as the difference between measured and PHSD results. The PHSD
model~\cite{PHSD} includes directly the dynamics of quark-gluon
degrees of freedom which are becoming more important with increasing
energy. We recall that the growing importance of the repulsive
partonic mean field --- illustrated earlier by the rise of the
elliptic flow explained convincingly in the PHSD
model~\cite{KBCTV11,KBCTV12} --- results here in an overestimation of
the CME background.

\begin{figure}[bht]
\includegraphics[width=0.45\textwidth,clip]{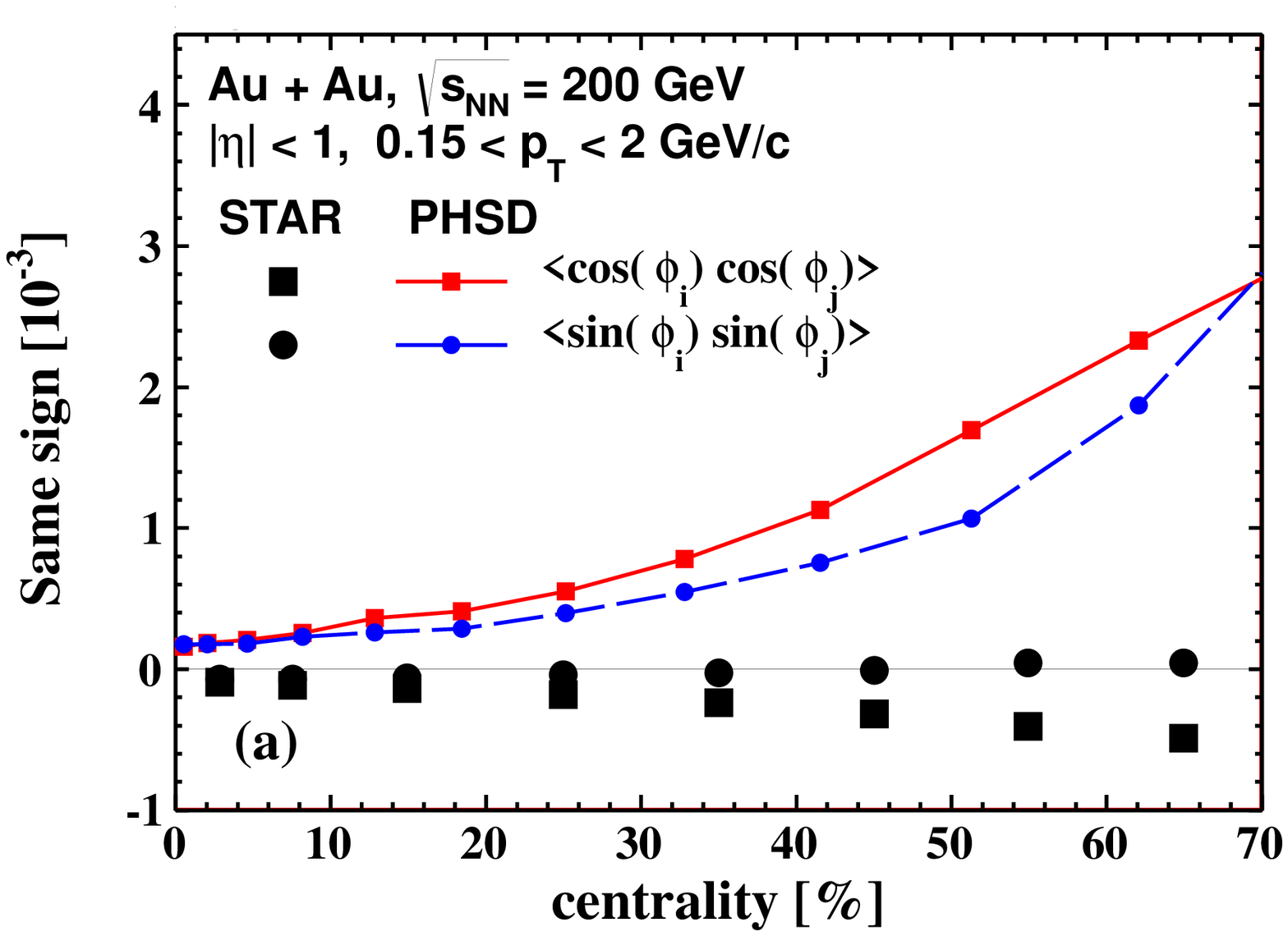}
\includegraphics[width=0.45\textwidth,clip]{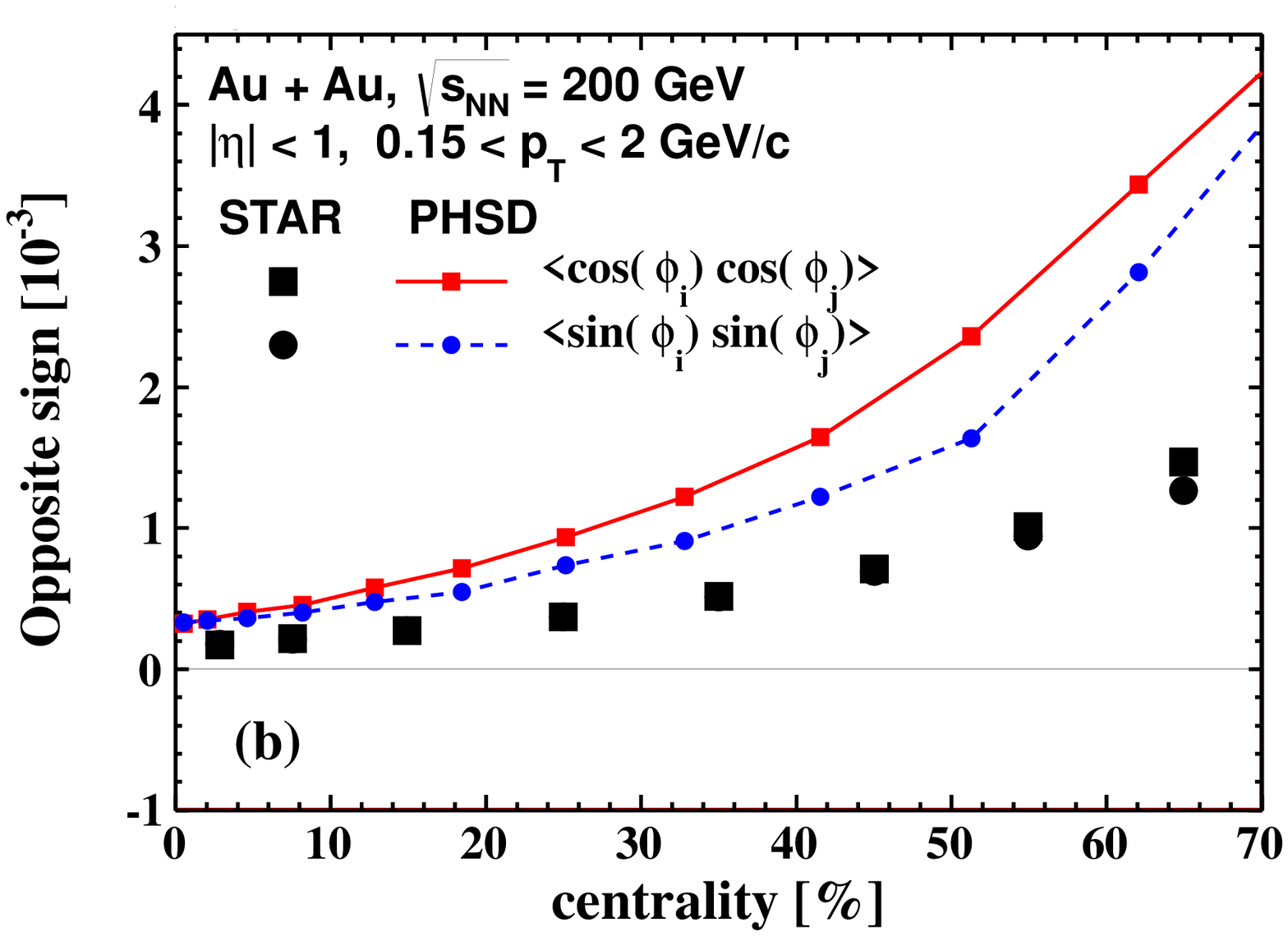}
\caption{(Color online) Angular correlations of $\cos$ (out-of-plane) and
  $\sin$ (in-plane) projections for Au+Au collisions at $\sqrt{s_{NN}}=$
  200 GeV as a function of centrality. The experimental data points
  are from~\cite{STAR-CME}.}
\label{CMEproj}
\end{figure}

In Fig.~\ref{CMEm} the results for the average cosine of the
difference in the azimuthal angles $\delta_{ij}$ are presented. The
measured centrality dependence for the same charge pions is flat and
practically consistent with zero while that for oppositely charged
particles is a monotonic increasing function with impact parameter.
However, the PHSD calculations clearly overestimate the experimental
points~\cite{STAR-CME}. We note in passing that the PHSD results for
Au+Au collisions at the energy $\sqrt{s_{NN}}=$ 200 GeV turn out to be
astonishingly close to the appropriate experimental data at
$\sqrt{s_{NN}}=$ 2.76 TeV~\cite{Ch11,LHC_CS12}. This fact indicates
that the strength of the repulsive scalar quark potential in PHSD
might be presently overestimated.

In accordance with Eq.~(\ref{cos}), one can separate the in-plane and
out-of-plane components using experimental results for $\gamma_{ij}$
and $\delta_{ij}$. Such separation together with PHSD calculation
results is presented in Fig.~\ref{CMEproj} for the same charge and
opposite charge pions. As was first noted in Ref.~\cite{BKL09} and is
seen in Fig.~\ref{CMEproj}(a), for the same-charge pairs the sinus
term is essentially zero whereas the cosine term is finite. This tells
us that the observed correlations are actually in-plane rather than
out-of-plane. This is contrary to the expectation from the chiral
magnetic effect, which results in same-charge correlations out of
plane. In addition, since the cosine term is negative, the in-plane
correlations are stronger for back-to-back pairs than for small angle
pairs. The PHSD does not reproduce these features. We see also that
for opposite-charge pairs the in-plane and out-of-plane correlations
are virtually identical. As was stated in~\cite{BKL09}, this is
difficult to comprehend since there is a sizable elliptic flow in
these collisions. Nevertheless, the PHSD model predicts very close
in-plane and out-of-plane distributions for opposite-charge pairs due
to scalar parton potential and at the same time nicely reproduces the
various harmonics of charged particles~\cite{KBCTV11,KBCTV12}. This
feature is not reproduced in the HSD.

We close this section with some more general remarks. As follows from
the results presented in Figs.~\ref{C2Hpr},\ref{CMEm},\ref{CMEproj} an
additional sizable source of asymmetry is needed for both in-plane and
out-of-plane components rather than only an out-of-plane component as
expected from the CME. As discussed in the Introduction, the vacuum
nontrivial topological structure (as a genuine source of the CME)
leads to the picture of a topological $\theta$ vacuum of non-Abelian
gauge theories. The $\theta$ term in the QCD Lagrangian explicitly
breaks ${\cal P}$ and ${\cal CP}$ symmetries of QCD. However,
stringent limits on the value of $\theta < 3\times 10^{-10}$ deduced
from the experimental bounds on the electric dipole moment of the
neutron~\cite{Ba06} practically indicate the absence of {\it global}
${\cal P}$ and ${\cal CP}$ violation in QCD. Reference to the {\it
  local} ${\cal P}$- and ${\cal CP}$-odd effects due to the
topological fluctuations characterized by an effective $\theta\equiv
\theta({\bf x},t)$ varying in space and time~\cite{Kharzeev:2004ey}
does not provide much hope. In addition, partons near the phase
transition are not chiral (as typically assumed) but massive degrees
of freedom in the PHSD in agreement with lattice QCD calculations. The
finite mass of the partons washes out the chirality effect.

\section{Summary and outlook}
\label{Sec:Summary}

In this study we have investigated several effects that might
contribute to the observed chiral magnetic effect (CME) in
relativistic nucleus-nucleus collisions on the basis of event-by-event
calculations within the PHSD transport approach. The individual
results can be summarized as follows:

\begin{itemize}

\item{Our study shows that fluctuations in the position of
  quasiparticles can manifest themselves in different interaction
  stages and in different ways.  Since the electromagnetic field
  generated by {\it spectators} is dominant at the early stage, the
  fluctuation in their position results in a noticeable fluctuation in
  the strength of the electromagnetic field. However, the fluctuation
  spread is not so large as expected in the estimate from
  Ref.~\cite{BS11} and its influence on observables is negligible; in
  particular, the event plane angle is not tilted due to these
  electromagnetic field fluctuations!}

\item{Early time fluctuations in the position of {\it participant}
  baryons were discussed in the past as a source of the impact
  parameter fluctuation. Its influence survives till the freeze-out
  resulting in a considerable difference between the theoretical
  reaction plane and the measured event plane. This effect leads to an
  increase in the magnitude of the elliptic flow and generates
  nonvanishing odd flow harmonics.}

\item{We have found out that within the PHSD model the retarded
  strong electromagnetic field --- created during nucleus-nucleus
  collisions --- turns out to be not so important as has been expected
  before.  Similarly to the HSD results in Ref.~\cite{EM_HSD}, the
  electromagnetic field has almost no influence on observables. The
  reason is not a shortness of the interaction time, when the
  electromagnetic field is maximal, but the compensation of the mutual
  action of transverse electric and magnetic components. This
  compensation effect might be important, for example, if an
  additional induced electric field (as a source of the CME) is
  available in the system since this field will not be entangled due
  to other electromagnetic sources.}

\item{Another important point emerging from the compensation effect
  of electric and magnetic forces is worth mentioning: A significance
  of an external magnetic field in astrophysics is largely accepted.
  There are many studies where various effects of external magnetic
  fields are discussed in the application to astrophysics ({\it e.g.},
  see the Introduction in Ref.~\cite{EM_HSD} and references in
  \cite{GMS11}. It is correct in this particular problems, however, in
  many cases it is concluded by a statement like ``the same effect
  should be observed in high-energy heavy-ion collisions'' which does
  not hold true due the compensation effect as demonstrated in the
  present work.}

\item{In the intermediate stage of the heavy-ion collision the
  statistical fluctuations of charged quasiparticles in momentum space
  can generate charge dipoles or even charge quadrupoles. However, the
  magnitudes $Q_{c1}$ and $Q_{c2}$ are small; their orientation is
  distributed almost uniformly and the direction of the main axis is
  changed from event to event. The influence of the electromagnetic
  field here is negligible again.}

\item{The transverse momentum conservation --- proposed as an
  alternative mechanism for an explanation of the observed azimuthal
  asymmetry --- shows a correlation of the CME and the elliptic flow.
  However, the effect estimated at $\sqrt{s_{NN}}=$ 200 GeV is too
  small and insensitive to the charge separation.}

\item{A possible charge separation of balancing charges has been
  addressed by the charge balance function. We note that the PHSD
  model fails to describe the focusing effect of the balance function
  for central Au+Au collisions. Certainly, further investigations of
  this problem are needed, both in theory and experiment especially at
  lower energies.}

\end{itemize}

The PHSD approach naturally takes into account the main alternative
mechanisms of the CME: the momentum conservation and local charge
conservation as well as clusters (mini jets, strings, prehadrons,
resonances). At the moderate energies $\sqrt{s_{NN}}=$ 7.7 and 11.5
GeV the PHSD model results are close to the experiment since partonic
degrees-of-freedom are subleading. However, at higher collision energy
the PHSD model fails to reproduce the observed azimuthal asymmetry. In
contrast with our earlier analysis within the HSD
model~\cite{BES-HSD}, the PHSD overestimates the measured centrality
dependence of azimuthal distributions due to an increasing action of
the repulsive scalar parton potential which generates the collective
flow harmonics in accordance with experiment. This finding suggests
that a new source of azimuthal anisotropy fluctuation is needed beyond
the `standard' interactions incorporated in PHSD. The new source does
not dominate in out-of-plane direction as could be expected for the
CME but both in-plane and out-of-plane components contribute with a
comparable strength. In this respect the interpretation of the CME
STAR measurements is still puzzling.

The present PHSD model is already quite elaborated, however, as our
analysis has shown, color degrees of freedom or intimate peculiarities
of non-Abelian Yang-Mills theory should additionally be taken into
consideration. In particular, this concerns the very early stage of
the nuclear interaction. In this initial state the highly compressed
strongly interacting matter is dense and though the QCD coupling
constant is small, gluonic states have high occupation numbers, {\it
  i.e.}, the partons begin to overlap in phase space which leads to
some saturated state. Strong color forces might create strong
chromoelectric and chromomagnetic fields producing a new state, a {\it
  glasma}~\cite{Fu12,LMcL06,La11} as mentioned above in context of the
discussion in Sec.~\ref{glasma}, or forming new objects like ``string
ropes'' described in the framework of Yang-Mills
theory~\cite{MCS02}. We are planning to include these effects into the
PHSD model in near future.

Another class of strong fields relevant to the chirality and
confinement of dynamical quarks is the long range (or soft) vacuum
gluon field configurations. Long range vacuum gluon fields can be seen
as an origin of nonzero gluon condensate and topological
susceptibility of QCD vacuum
\cite{Shifman:1978bx,Minkowski:1981ma}. Soft fields arise in the
consideration of the global minima of the QCD effective
action\cite{Leutwyler:1980ev} and are known to play an important role
in hadron phenomenology at zero temperature \cite{Kalloniatis:2003sa}.
The nonzero gluon condensate survives at high temperature as
demonstrated by QCD lattice calculations \cite{D'Elia:2002ck}.
Interplay of strong electromagnetic and vacuum long-range gluon fields
can lead to the qualitatively new effects in high energy heavy ion
collisions \cite{Galilo:2011nh}. However these effects are beyond the
scope of this paper.

\section*{ACKNOWLEDGMENTS}

We are thankful to Che Ming Ko, Sergei Molodtsov, Sergei Nedelko, Oleg
Teryaev, Sergei Voloshin, and Harmen Warringa for illuminating
discussions. This work has been supported by the LOEWE Center HIC for
FAIR, a Heisenberg-Landau grant, RFFI grant no.\ 11-02-01538-a, and by
DFG.

\end{document}